\documentclass[a4paper,usenames,dvipsnames,11pt]{article}
\pdfoutput=1

\usepackage{jheppub}
\usepackage{slashed}
\usepackage{mathrsfs,booktabs,multirow,tabularx}
\usepackage{stmaryrd}
\usepackage{xspace}
\usepackage{fancyvrb}
\usepackage[makeroom]{cancel}
\usepackage{amsmath}    
\usepackage{amssymb}    %
\usepackage{graphicx}   
\usepackage{verbatim}   
\usepackage{lscape}
\usepackage{subfig}
\usepackage{listings}
\usepackage{hyperref}

\renewcommand\arraystretch{1.1}

\def\beq{\begin{equation}}
\def\eeq{\end{equation}}
\def\beqn{\begin{eqnarray}}
\def\eeqn{\end{eqnarray}}

\newcommand\sss{\scriptscriptstyle}

\newcommand\ep{\epsilon}

\newcommand\half{\frac{1}{2}}

\newcommand\as{\alpha_{\sss S}}

\newcommand\aem{\alpha}
\newcommand\eff{\varepsilon}
\newcommand\ord{{\cal O}}

\newcommand\GenMC{{\cal F}_{\mbox{\tiny MC}}}
\newcommand\GenNLO{{\cal F}_{\mbox{\tiny MC@NLO}}}
\newcommand\GenNLOD{{\cal F}_{\mbox{\tiny MC@NLO-$\Delta$}}}
\newcommand\aNLO{{\sc\small MadGraph5\_aMC@NLO}}
\newcommand\aNLOs{{\sc\small MG5\_aMC}}
\newcommand\aNLOD{{\mbox{{\rm MC@NLO}-$\Delta$}}}
\newcommand\PYe{{\sc\small Pythia8}}
\newcommand\clH{{\mathbb H}}
\newcommand\clS{{\mathbb S}}
\newcommand\conf{{\cal K}}
\newcommand\confH{\conf^{(\clH)}}
\newcommand\confS{\conf^{(\clS)}}
\newcommand\sigmaH{\sigma^{\sss (\clH)}}
\newcommand\sigmaS{\sigma^{\sss (\clS)}}
\newcommand\sigmaHD{\sigma^{\sss (\Delta,\clH)}}
\newcommand\sigmaSD{\sigma^{\sss (\Delta,\clS)}}
\newcommand\sigmaNLO{\sigma^{\sss\rm (NLO)}}
\newcommand\sigmaNLOa{\sigma^{{\sss (\rm NLO},\alpha{\sss )}}}
\newcommand\sigmaNLOE{\sigma^{\sss {(\rm NLO},E)}}
\newcommand\sigmaMC{\sigma^{\sss\rm (MC)}}

\newcommand\phsp{d\phi}

\newcommand\phspnpo{\phsp_{n+1}}
\newcommand\Phsp{\Phi}
\newcommand\Phspn{\Phsp^{(n)}}
\newcommand\Phspnpo{\Phsp^{(n+1)}}
\newcommand{\meas}{\chi}
\newcommand{\nclo}{{\bf N.1}}
\newcommand{\nclt}{{\bf N.2}}
\newcommand{\nclth}{{\bf N.3}}
\newcommand\ident{{\cal I}}

\newcommand\stepf{\Theta}

\newcommand{\pT}{p_{\sss T}}
\newcommand{\pt}{p_{\sss T}}
\newcommand{\kt}{k_{\sss T}}
\newcommand{\Ht}{H_{\sss T}}
\newcommand{\qt}{q^2}
\newcommand{\qmt}{q_{\min}^2}
\newcommand{\Qt}{Q^2}

\newcommand{\Qot}{Q_1^2}
\newcommand{\Qtt}{Q_2^2}
\newcommand{\QMt}{Q_{\max}^2}

\newcommand{\hP}{\hat{P}}
\newcommand\Nell{N_\ell}
\newcommand\Mdips{{\mathbb M}}
\newcommand\mdip{m_{dip}}
\newcommand\Sfun{{\cal S}}
\newcommand\bK{\bar{K}}
\newcommand\bk{\bar{k}}
\newcommand\bp{\bar{p}}

\newcommand\ccol{{\cal C}}
\newcommand\mother{\overline{i\oplus j}}
\newcommand\II{{\tt II}}
\newcommand\IF{{\tt IF}}
\newcommand\FI{{\tt FI}}
\newcommand\FF{{\tt FF}}
\newcommand\CF{C_{\sss F}}
\newcommand\CA{C_{\sss A}}
\newcommand\bt{\bar{t}}
\newcommand\bb{\bar{b}}
\newcommand\bq{\bar{q}}
\newcommand{\gev}{\,\textrm{GeV}}

\title{On the reduction of negative weights in MC@NLO-type matching
procedures}

\author[a]{R. Frederix,}
\author[b]{S. Frixione,}
\author[a]{S. Prestel,}
\author[c]{P. Torrielli}
\affiliation[a]{Theoretical Particle Physics, Department of Astronomy 
and Theoretical Physics, Lund University, S\"olvegatan\ 14\ A, SE-223\ 62 
Lund, Sweden}
\affiliation[b]{INFN, Sezione di Genova, Via Dodecaneso 33, I-16146, 
Genoa, Italy}
\affiliation[c]{Dipartimento di Fisica and Arnold-Regge Center, 
Universit\`a di Torino and INFN, Sezione di Torino, Via P. Giuria 1, 
I-10125, Turin, Italy}

\emailAdd{rikkert.frederix@thep.lu.se}
\emailAdd{Stefano.Frixione@cern.ch}
\emailAdd{stefan.prestel@thep.lu.se}
\emailAdd{torriell@to.infn.it}

\abstract{We show how a careful analysis of the behaviour of a
parton shower Monte Carlo in the vicinity of the soft and collinear
regions allows one to formulate a modified MC@NLO-matching prescription 
that reduces the number of negative-weight events with respect to that 
stemming from the standard MC@NLO procedure. As a first practical
application of such a prescription, that we dub MC@NLO-$\Delta$, 
we have implemented it in the {\sc\small MadGraph5\_aMC@NLO} 
framework, by employing the {\sc\small Pythia8} Monte Carlo. 
We present selected MC@NLO-$\Delta$ results at the 13~TeV LHC, 
and compare them with MC@NLO ones. We find that the former predictions 
are consistent with the latter ones within the typical matching systematics, 
and that the reduction of negative-weight events is significant.
}

\keywords{QCD, NLO matching, Parton Shower Monte Carlos}

\preprint{
\begin{flushright}
LU-TP~20-09\\
\today
\end{flushright}
}

\begin{document}
\maketitle
\flushbottom

\section{Introduction\label{sec:intro}}
\begin{figure}[htb]
\begin{center}
  \includegraphics[width=0.65\textwidth]{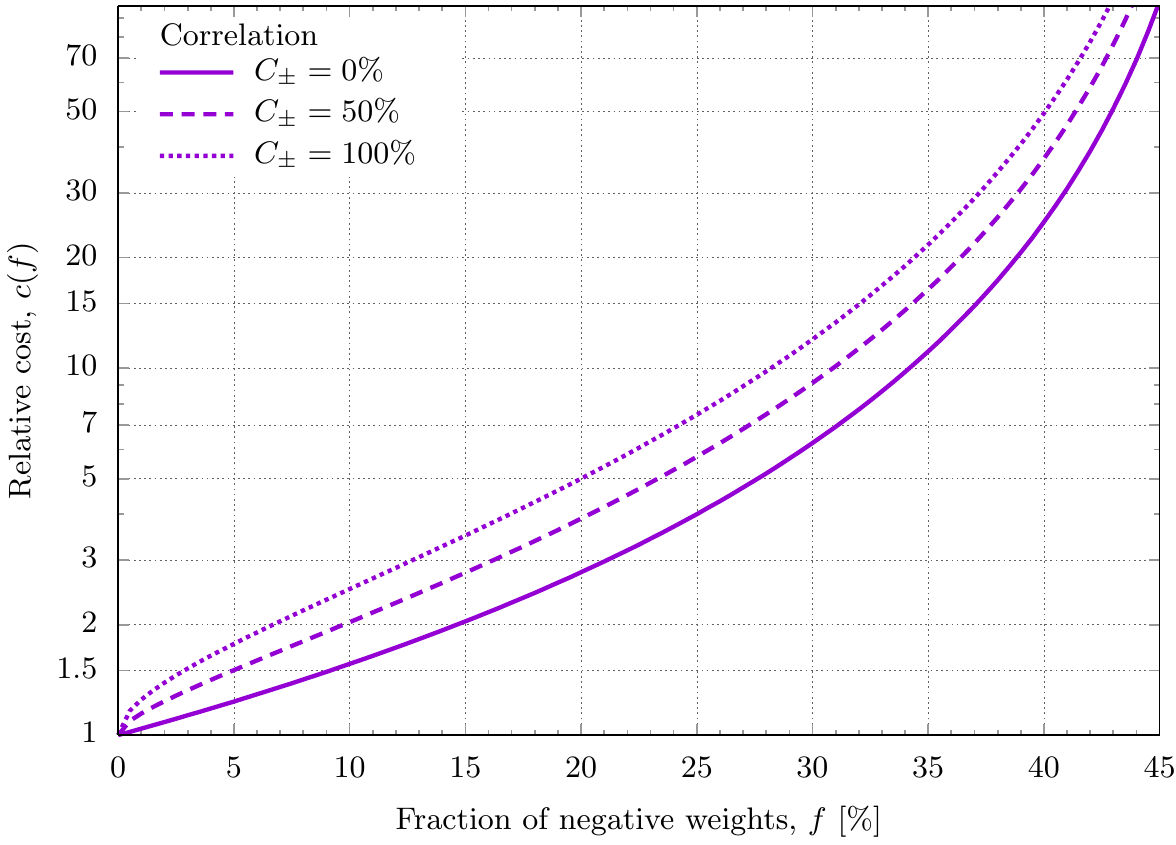}
\end{center}
\caption{\label{fig:cost} 
Relative cost as a function of the fraction of negative weights,
eq.~(\ref{costdef}), for three different values of the correlation
parameter $C_{\pm}$.
}
\end{figure}

It is inconceivable that modern experimental high-energy particle
physics be done without the massive use of event generators, which
are also enjoying an increasing number of applications in theoretical
phenomenology. This has stimulated a vigorous research activity in 
the past two decades, the upshot of which is that event generators,
while maintaining their traditional flexibility, have significantly
increased their predictive power (and its most important spinoff,
the reduction of systematics), through matching and merging with
perturbative matrix-element computations. Apart from a few selected
cases, matching and merging are either carried out at the tree level
or at the next-to-leading order (NLO). In this work, we shall consider only 
the latter simulations, that are more accurate than the former, but require
more involved computations (which nowadays can fortunately be automated).
On top of that, one must bear in mind that NLO cross sections are not 
positive-definite locally in the phase space. This implies that some of the 
hard events that will eventually be showered have negative weights. It is 
convenient to introduce an efficiency associated with the fraction of 
negative weights; denoting the latter by $f$, such an efficiency is 
defined as follows:
\beq
\eff(f)=1-2f\,,
\;\;\;\;\;\;\;\;
0\le f<0.5
\;\;\;\;\;\;\Longrightarrow\;\;\;\;\;\;
0<\eff(f)\le 1\,.
\label{effdef}
\eeq
This efficiency stems from matrix-element computations 
(see e.g.~eq.~(4.33) of ref.~\cite{Frixione:2002ik}), but it
affects the overall performance of the latter only insofar that
it is correlated with other efficiencies relevant to them; for
example, it is usually the case that phase-space integration
converges slightly faster in the case of a positive-definite
integrand than in the case of an integrand whose sign can change;
similar considerations are valid e.g.~for the unweighting efficiency.
However, for all practical purposes such correlations can be neglected:
by far, the most significant impact of $\eff(f)$ being smaller than one
is that it implies that, in order to obtain the same statistical accuracy 
at the level of physical observables as the one of a positive-definite
simulation, a number of events must be generated which is larger
than in the latter case. In order to quantify this statement, let us
denote by:
\beq
N\,,
\;\;\;\;\;\;\;\;
N_+=(1-f)N\,,
\;\;\;\;\;\;\;\;
N_-=fN
\;\;\;\;\;\;\Longrightarrow\;\;\;\;\;\;
N=N_+ + N_-\,,
\label{NNpNm}
\eeq
the total number of events, the number of events with positive weights,
and the number of events with negative weights, respectively. Furthermore,
we assume such events to be unweighted\footnote{Discussing efficiencies
becomes significantly more complicated in the case of weighted events.
Of course, this is not the (main) reason why unweighted events must be 
preferred to weighted events whenever possible: rather, they constitute 
a much more realistic representation of actual physical events, and their
samples are much smaller, for any given accuracy target, than those relevant
to weighted events.}: in other words, their weights are equal to $\pm\omega$, 
with $\omega>0$ a factor that is constant for all events in a given 
generation and that includes the suitable normalisation. The resulting 
cross section and its associated error will thus be:
\beqn
\sigma&=&\omega\left(
N_+ - N_- \pm \sqrt{N_+ + N_- +2C_{\pm}\sqrt{N_+}\sqrt{N_-}}\right)
\nonumber
\\*&=&
\omega\left(\eff(f)N\pm\sqrt{1+C_{\pm}\sqrt{1-\eff(f)^2}}\sqrt{N}\right),
\label{sigpm}
\eeqn
where by $C_{\pm}$ we have denoted the non-negative correlation between 
positive- and negative-weight events (this number is typically neglected).
In the context of a positive-definite simulation, where $M$ 
unweighted events are generated with weights all equal to $\omega^\prime>0$,
the analogue of eq.~(\ref{sigpm}) reads:
\beq
\sigma=\omega^\prime\left(M\pm\sqrt{M}\right).
\label{sigpos}
\eeq
By imposing the cross sections in eqs.~(\ref{sigpm}) and~(\ref{sigpos}) 
to have the same relative error we obtain:
\beq
N=c(f)\,M\,,
\label{NvsM}
\eeq
where
\beq
c(f)=\frac{1+C_{\pm}\sqrt{1-\eff(f)^2}}{\eff(f)^2}\,.
\label{costdef}
\eeq
The fact that $N\ge M$ (with $N=M$ if and only if $f=0$) formalises 
what was stated before eq.~(\ref{NNpNm}): a simulation that features events 
of either sign can attain the same statistical accuracy as one that is 
positive definite only by generating a number of events which is larger 
by a factor $c(f)$ w.r.t.~that of the latter. For this reason, we call 
$c(f)$ the {\em relative cost} of the simulation; we plot this quantity 
as a function of $f$ in fig.~\ref{fig:cost}, for $C_{\pm}=0, 0.5, 1$.

The discussion thus far has been rather schematic. For example, we have
implicitly assumed $f$ to be independent of the kinematics, which is 
never the case. More correctly, one would need to use eqs.~(\ref{sigpm}) 
and~(\ref{sigpos}) locally in the phase space, thus defining a local
relative cost, and subsequently construct the global relative
cost as the weighted (by number of events) average of the local ones.
In practice, eq.~(\ref{costdef}), with $f$ the overall fraction of 
negative-weight events, does characterise well enough the behaviour
of simulations with events of either sign, and we shall often use it
in the following.

The problem with $c(f)>1$ for any $f>0$ is not statistics {\em per se},
but the fact that it generally implies additional {\em financial} costs:
longer running times, hence larger power consumption (events with 
negative weights contribute to climate change!), and bigger storage
space, to name just the most important ones. Denoting by $p$ ($p^\prime$)
the overall price tag for the generation, full simulation, analysis, 
and storage of an individual event resulting from a positive-definite
(non-positive-definite) simulation, the additional costs alluded to 
before are:
\beq
Np^\prime-Mp=M\,\big[c(f)p^\prime-p\big]\,.
\label{addcost}
\eeq
With all other things being equal (and chiefly among them, the control
of the theoretical systematics), it is therefore advantageous to make
$f$ as small as possible, so as to minimise the additional 
costs\footnote{Note that $p^\prime-p$ can have either sign, although
when NLO and LO calculations are taken as examples of non-positive- and
positive-definite simulations, respectively, most likely $p^\prime>p$.
In any case, in the context of a complete experimental analysis the
contribution to the cost due to the generation phase alone is minor, 
and thus $p^\prime\simeq p$.}
of eq.~(\ref{addcost}). This is the goal of the present work, in the
context of the MC@NLO matching formalism~\cite{Frixione:2002ik}.

Before proceeding, we remind the reader that currently the vast majority
of theoretical studies, and essentially all of the NLO$+$parton shower 
simulations performed 
by experimental collaborations, are based on either the MC@NLO or 
the POWHEG~\cite{Nason:2004rx} methods, and on their closely-related
variants~\cite{Hoeche:2011fd,Platzer:2011bc}; alternative approaches
(see e.g.~refs.~\cite{Nagy:2005aa,Bauer:2006mk,Nagy:2007ty,
Giele:2007di,Bauer:2008qh,Jadach:2015mza}) are much less
well-developed, and rarely used in practice. At variance with MC@NLO,
the POWHEG formalism is characterised by a very small fraction of negative 
weights\footnote{This fraction is not necessarily equal to zero, and 
depending on the running conditions is not necessarily small -- see 
e.g.~ref.~\cite{Oleari:2011ey}.}. Such an appealing feature relies, among
other things, on the exponentiation of real-emission matrix elements.
This induces a few characteristic features, in particular because
neither these matrix elements nor their associated phase spaces 
exponentiate away from the soft and collinear regions. If we denote 
by $\as^b$ the Born-level perturbative order relevant to a given computation, 
at $\ord(\as^{b+2})$ (and beyond) the POWHEG physical prediction will 
contain potentially large terms whose origin is due to matrix elements of 
$\ord(\as^{b+1})$, in addition to those naturally resulting from Monte-Carlo 
(MC henceforth) parton showers. This is avoided by construction in 
the MC@NLO matching, where all terms of $\ord(\as^{b+2})$ and beyond are 
solely of MC origin\footnote{A more precise version of these statements is 
the following: if no showers are performed, terms of order higher than 
$\as^{b+1}$ are (are not) equal to zero in an MC@NLO (an exclusive POWHEG) 
prediction. Note, also, that we are not implying that higher-order terms 
of MC origin are necessarily ``small'' or ``correct'' -- however, 
the behaviour of the MC is a given for matching techniques, and addressing 
it is beyond the scope of the latter procedures.\label{ft:extrat}}.

The central idea of the present work is that of showing that, by a
suitable modification of the $\ord(\as^{k})$ terms ($k\ge 2$) in an 
MC@NLO cross section by means of contributions of sole MC origin,
one arrives at a matching prescription, that 
we call $\aNLOD$, which preserves the good features of MC@NLO while
significantly reducing the fraction of negative weights. In particular,
the formal $\ord(\as^{b+1})$ expansion of the $\aNLOD$ cross section 
coincides with that of MC@NLO and hence, by construction, with the 
NLO result. In the infrared regions, the logarithmic behaviour of $\aNLOD$ is
the same as that of the MC one matches to, as is the case for MC@NLO.
Furthermore, the $\aNLOD$ formulation leads in a natural manner to
a richer structure of the hard events to be showered by the MC. Namely, 
a (generally different) shower scale is associated with each 
oriented colour line, and is passed to the MC; this is in contrast 
to what is done currently, where a single scale characterises the production
process, and is regarded as a global property of the event.
Finally, we note that in NLO-matched MCs the fraction of negative-weight 
events with Born kinematics can be reduced by means of a procedure that has 
been called {\em folding} in ref.~\cite{Nason:2007vt}, and implemented in 
practice in POWHEG in ref.~\cite{Alioli:2010xd}\footnote{In MC@NLO, this 
possibility had been envisaged in the original paper~\cite{Frixione:2002ik}
(see the bottom of sect.~4.5 there), but not implemented in computer codes 
so far.}. Here we shall show that, on top of the intrinsic reduction of 
the number of negative-weight events in the $\aNLOD$ matching, the folding 
technique is rather effective there, whereby such a reduction is more 
pronounced than in the case of the standard MC@NLO matching. 

This paper is organised as follows. In sect.~\ref{sec:negw} we review
the basic properties of the MC@NLO matching that lead to negative-weight
events, and classify the latter. In sect.~\ref{sec:mcnloD} we introduce
the new matching prescription, $\aNLOD$, that constitutes the core of
this work. This is based on a scalar quantity, $\Delta$, whose construction
we describe in detail in sect.~\ref{sec:Dcostr}; the hard-event multi-scale 
structure it induces is introduced in sect.~\ref{sec:LHEscales}. Some 
peculiar features of the practical implementation of $\aNLOD$ are 
reported in sect.~\ref{sec:impl}. In sect.~\ref{sec:res} we present 
sample hadroproduction $\aNLOD$ results, which we systematically 
compare with their MC@NLO counterparts. We draw our conclusions 
in sect.~\ref{sec:conc}. Finally, some technical information on
the \PYe\ Sudakov form factors are collected in appendix~\ref{sec:PYesud}.

\section{Anatomy of negative weights in MC@NLO\label{sec:negw}}
We start by pointing out that in MC@NLO the exact amount of negative-weight
events and their distribution in the phase space depend on several elements.
Among these, the most important are the following: the parton shower
MC one matches to, and in particular its shower variables; the technique,
which is typically a subtraction procedure, used to compute the
underlying NLO cross section; and the choice of the phase-space
parametrisation employed in the latter computation. Therefore, in
a bottom-up approach to the reduction of negative weights, one 
would {\em construct} the shower and the short-distance computations
with the specific goal of minimising $f$. This is a potentially very
interesting strategy, which however appears to be quite complicated; 
we shall not pursue it here. Rather, we shall follow a top-down approach,
where both the shower and the NLO calculations are considered as given,
and it is the matching between them which is responsible for the 
minimisation of $f$. This can be done thanks to the fact that, in spite 
of their specific features mentioned above, negative weights possess 
universal characteristics which one can exploit to reduce their number.
In order to discuss such universal characteristics, we now sketch out the
basic MC@NLO formulae, simplifying them as much as possible, lest the 
details obscure the basic ideas. If the reader wants to be definite,
explicit expressions based on FKS subtraction~\cite{Frixione:1995ms,
Frixione:1997np} can be found e.g.~in refs.~\cite{Frederix:2009yq,
Alwall:2014hca,Frederix:2018nkq} for the \aNLO\ implementation (\aNLOs\ 
henceforth).

The key simplification from a notational viewpoint stems from one of the 
basic features of the FKS subtraction and of the MC@NLO implementations
based on it. Namely, for any given real-emission process the phase space
is partitioned in an effective manner by means of the $\Sfun$ functions,
so that one ultimately deals with a linear combination of short-distance
cross sections which have, at most, one soft and one collinear singularity.
Such a partition singles out two partons, called the FKS parton and its
sister, with which the soft and collinear singularities are associated. 
We shall thus work by using the rule:
\begin{itemize}
\item[{\em R.1}:] The following formulae assume that the real-emission
process, the FKS parton (labelled by $i$), and the sister of the latter
(labelled by $j$) are given and fixed. In order to obtain the physical
cross sections, one must sum over these quantities.
\end{itemize}

\noindent
Bearing the above condition in mind, the MC@NLO generating functional 
is written as follows\footnote{As an example of the simplifications
induced by rule {\em R.1}, the reader is encouraged to compare 
eq.~(\ref{genMCatNLO}) with eq.~(2.121) of ref.~\cite{Alwall:2014hca}.}:
\beq
\GenNLO=\GenMC\!\left(\confH\right)d\sigmaH
+\GenMC\!\left(\confS\right)d\sigmaS\,,
\label{genMCatNLO}
\eeq
where $\GenMC$ is the generating functional of the MC one matches to.
By $\confH$ and $\confS$ we have denoted $\clH$- and $\clS$-event
kinematic configurations, respectively. For example, if Born-level
processes for the cross section of interest feature $n$ final-state
particles, $\confH$ and $\confS$ correspond to \mbox{$2\to n+1$} and
\mbox{$2\to n$} configurations, respectively\footnote{In order to simplify
the notation, we assume here that all of the particles relevant to our 
processes are strongly interacting. It is easy to include {\em a posteriori} 
extra particles which are not strongly interacting.}. The short-distance 
cross sections on the r.h.s.~of eq.~(\ref{genMCatNLO}) are:
\beqn
d\sigmaH&=&d\sigmaNLOE-d\sigmaMC\,,
\label{xsecH}
\\
d\sigmaS&=&d\sigmaMC+
\sum_{\alpha=S,C,SC} d\sigmaNLOa\,.
\label{xsecS}
\eeqn
Here, we have denoted by $d\sigmaMC$ the MC counterterms; the other
contributions are identical to those that enter an NLO {\em fixed-order}
cross section:
\beq
\frac{d\sigmaNLO}{d\conf}=\delta\!\left(\conf-\confH\right)d\sigmaNLOE+
\delta\!\left(\conf-\confS\right)\sum_{\alpha=S,C,SC} d\sigmaNLOa\,.
\eeq
Thus, $d\sigmaNLOE$ is the real-emission contribution, while
$d\sigmaNLOa$, \mbox{$\alpha=S,C,SC$} collect all of the other terms
(the Born, and contributions of virtual, soft, collinear, and soft-collinear 
origin; in a non-FKS language, the latter are therefore the integrated 
and unintegrated fixed-order counterterms). We point out that 
the cross sections on the r.h.s.~of eqs.~(\ref{xsecH}) and~(\ref{xsecS})
have support in an $(n+1)$-body phase space. We write the latter as
follows\footnote{Throughout this paper, we understand the integration over
Bjorken $x$'s and the definition of the integration variables associated 
with them.}:
\beq
\phspnpo=\Phspnpo\left(\meas_{n+1}\right)\,d\meas_{n+1}\,,
\label{phspnpo}
\eeq
where $\meas_{n+1}$ denotes the set of the chosen \mbox{$3n-1$}
integration variables, whose nature need not be specified here,
except for the fact that its has the following general form:
\beqn
\meas_{n+1}&=&\meas_n\bigcup\meas_r\,,
\label{intvnpo}
\\
\meas_r&=&\big\{\xi,y,\varphi\big\}\,.
\label{radvar}
\eeqn
By $\meas_n$ we have denoted \mbox{$3n-4$} integration variables
that define $n$-body (i.e.~Born-level) configurations, and by $\meas_r$ 
the variables that parametrise the extra radiation that occurs at the
real-emission level. In an FKS framework (where one works in the 
c.m.~frame of the incoming partons), $\xi$ is the rescaled FKS-parton 
energy, and $y$ the cosine of the angle between the FKS parton
and its sister; $\varphi$ is an azimuthal angle. Thus, $\xi\to 0$ and
$y\to 1$ correspond to the soft and collinear limits, respectively.
One can always construct the phase spaces so that
(see e.g.~ref.~\cite{Frederix:2009yq}):
\beq
\confS=\confS(\meas_n)\,,
\;\;\;\;\;\;\;\;
\confH=\confH(\meas_{n+1})\equiv\confH(\meas_n,\xi,y,\varphi)\,,
\label{cf1}
\eeq
and
\beq
\confS(\meas_n)=
\confH(\meas_n,0,y,\varphi)=
\confH(\meas_n,\xi,1,\varphi)\,.
\label{KSvsH}
\eeq
Equation~(\ref{KSvsH}) gives an unambiguous meaning to the connection
between a real-emission configuration and its underlying Born-level
configuration. Furthermore, eqs.~(\ref{phspnpo})--(\ref{KSvsH}) imply: 
\beq
\phspnpo=J(\meas_n,\meas_r)\,\Phspn\left(\meas_n\right)\,d\meas_n d\meas_r\,,
\label{phspnporad}
\eeq
where $J$ is a factor, whose explicit form is not relevant here, of 
Jacobian origin. Finally, we define the pull:
\beq
P(\confH)\equiv P(\meas_n,\xi,y,\varphi)\,,
\label{pulldef}
\eeq
as a variable that measures the distance (in phase space) between a 
real-emission configuration and its underlying Born-level configuration. 
Therefore, the pull must be such that:
\beq
\lim_{\xi\to 0}P(\confH)=\lim_{y\to 1}P(\confH)=0\,.
\label{pulllim}
\eeq
For example, in Drell-Yan production $P$ can be identified with the 
transverse momentum of the lepton pair. We note that, for any given
process, there is ample freedom to define the pull. However, for the
sake of the present discussion its precise definition is irrelevant;
what matters is that, by assuming that $P$ has canonical dimensions
equal to one (which is not restrictive), and by denoting by $M_H$
the typical hard scale of the process, owing to eq.~(\ref{pulllim})
the regions:
\beqn
&&\phantom{aaaaaaaaa}
P(\confH)\ll M_H\,,
\label{pull1}
\\*
&&P(\confH)\sim M_H\;\;\;\;\bigcup\;\;\;\;
P(\confH)>M_H\,,
\label{pull2}
\eeqn
correspond to $\confH$ being a soft- and/or collinear-emission configuration, 
and an intermediate- or hard-emission configuration, respectively.

We classify negative-weight events in MC@NLO as follows:

\vskip 0.4truecm
\hskip -0.3truecm
\begin{minipage}{0.7\textwidth}
\begin{itemize}
\item[\nclo] $\clH$ events with $P(\confH)\ll M_H$;
\item[\nclt] $\clH$ events with $P(\confH)\sim M_H$;
\item[\nclth] $\clS$ events.
\end{itemize}
\end{minipage}

\vskip 0.5truecm
\noindent
Events of both classes \nclo\ and \nclt\ are due to the fact that
the MC counterterms might overestimate the real-emission cross section,
and thus the linear combination in eq.~(\ref{xsecH}) is negative. 
\nclo\ events will be cancelled after showering (i.e.~at the level of 
physical cross sections) by $\clS$ events; being in an MC-dominated region
and thanks to the fact that the number of $\clS$ events is generally
much larger than that of $\clH$ events, such a cancellation occurs with high 
efficiency\footnote{The presence of events of class \nclth\ just lowers 
this efficiency, but does not hamper the cancellation.}. By far and large,
this also implies that they affect very mildly the shape of kinematical 
distributions\footnote{As for all $\clH$ events in MC-dominated regions.}, 
their main impact being on the absolute normalisation (we remind the
reader that the MC@NLO and fixed-order NLO total cross sections, before
acceptance cuts, are identical). 

A similar cancellation mechanism is at play in the case of \nclt\ events.
However, since the region $P(\confH)\sim M_H$ is not dominated by MC
effects, the cancellation efficiency is lower than that relevant to
\nclo\ events. However, the number of \nclt\ events is smaller than 
that of \nclo\ events: the larger the pull, the smaller the probability 
that the MC will be able to generate the corresponding kinematic 
configuration. While this is always qualitatively true, quantitatively
it depends on the running conditions of the MC, and in particular on
the choice of the shower starting scale. We point out that the rationale
behind such a choice is different if one does not or does match the MC
to matrix elements. In the former case, larger scales are preferred,
in order to fill the phase space more effectively, at the cost of 
stretching the simplifying approximations upon which the MC is 
based\footnote{We understand ``larger scales'' here not in absolute 
value, but relative to the natural hard scale of the process.}.
This is not necessary in the latter case, since hard emissions are already 
provided by the matrix elements; thus, when performing a matching smaller 
shower starting scales are a more appealing option. Note that while this
conclusion is purely due to physics arguments, one of its by-products
is that it helps reduce the number of events of class \nclt.

If the existence of \nclo\ and \nclt\ events stems from the basic
physics of the MC@NLO cross section, that of \nclth\ events is 
predominantly due to the technical procedure that is (normally) used 
to generate such events. Note that $\clS$ events have an $n$-body 
kinematics (see the rightmost term on the r.h.s.~of eq.~(\ref{genMCatNLO})),
but their associated short distance cross section, eq.~(\ref{xsecS}),
has support in an $(n+1)$-body phase space. This implies that, if
weighted events are defined, their kinematic configurations depend
solely on the variables $\meas_n$, while their associated weights
depend on both $\meas_n$ and $\meas_r$ 
(see eqs.~(\ref{cf1})--(\ref{phspnporad})). In the case of 
unweighted-event generation, the hit-and-miss procedure is carried out 
in the $(n+1)$-body phase space. Therefore, the same $\meas_r$-independent
configuration $\confS$ may be obtained multiple times, depending on
the sampling of the $\meas_r$ subspace; this is a major source
of negative weights in $\clS$ events. On the other hand, it should 
be clear that the $\clS$-event generating functional can be rewritten 
as follows\footnote{The general idea that underpins eq.~(\ref{genSevI})
is that of the folding, that has been already alluded to in 
sect.~\ref{sec:intro}; see  sect.~\ref{sec:impl} for more details.}:
\beq
\GenMC\!\left(\confS\right)\int_{\meas_r}\!d\sigmaS\,.
\label{genSevI}
\eeq
In other words, by adopting eq.~(\ref{genSevI}) one first integrates the 
short-distance cross section $d\sigmaS$ over the $\meas_r$ subspace, and 
then generates (weighted or unweighted) events. This loss of locality in 
$\meas_r$, which has no consequence on physics since $\confS$ is independent
of $\meas_r$, is advantageous because it reduces the number of negative
weights. In fact, the integrated short-distance cross section in
eq.~(\ref{genSevI}), at variance with its un-integrated counterpart
$d\sigmaS$, is generally positive. This can be understood by a simple
perturbative argument: linear combinations of quantities that have the
same kinematic structure (which is strictly the case of the $\clS$-event
cross section upon integration over $\meas_r$) are dominated, in a 
well-behaved perturbative expansion, by the lowest-order terms, in this 
case the Born, which is positive-definite. To recapitulate the general 
argument that informs eq.~(\ref{genSevI}): a function ($d\sigmaS$ in this 
case) with support in a $(3n-1)$-dimensional space can be locally negative 
in such a space, but positive-definite in the $(3n-4)$-dimensional space 
obtained by integrating over three variables ($\meas_r$ in this case) of the 
former. Thus, if the unweighting of such a function is performed in the 
$(3n-1)$-dimensional ($(3n-4)$-dimensional) space, negative weights will 
(will not) occur.

In summary, the reduction of negative-weight $\clS$ events can be
achieved without changing the MC@NLO prescription, simply by means of
eq.~(\ref{genSevI}). Conversely, the case of $\clH$ events is more
involved, and requires a modification of the matching procedure, which
we shall detail in the next section. In the context of this new
prescription, that we call $\aNLOD$, the reduction of class-\nclth\
events can again by achieved thanks to the analogue of eq.~(\ref{genSevI}).

\section{The $\aNLOD$ matching prescription\label{sec:mcnloD}}
Consider the following generating functional:
\beq
\GenNLOD=\GenMC\!\left(\confH\right)d\sigmaHD
+\GenMC\!\left(\confS\right)d\sigmaSD\,,
\label{genMCatNLOD}
\eeq
where:
\beqn
d\sigmaHD&=&\big(d\sigmaNLOE-d\sigmaMC\big)\Delta\,,
\label{xsecHD}
\\
d\sigmaSD&=&d\sigmaMC\Delta+
\sum_{\alpha=S,C,SC} d\sigmaNLOa+
d\sigmaNLOE\big(1-\Delta\big)\,.
\label{xsecSD}
\eeqn
The quantity $\Delta$ is understood to have support in the $(n+1)$-body 
phase-space, and to obey the condition \mbox{$0\le\Delta\le 1$}.
Bearing in mind the discussion on both the characteristics of 
class-\nclo\ events and of all $\clH$ events in MC-dominated regions 
(see sect.~\ref{sec:negw}), we expect a reduction of the former ones with
negligible effects on the shapes of physical distributions in the latter 
regions if $\Delta$ is such that:
\beq
\Delta\;\longrightarrow\;0
\;\;\;\;\;\;\;\;\;\;
{\rm soft~and~collinear~limits}.
\label{Dcond1}
\eeq
Note, in fact, that with eq.~(\ref{Dcond1}) one manifestly obtains
\mbox{$\GenNLOD\propto \GenMC\!\left(\confS\right)$} in the soft and
collinear regions, analogously to what happens with the standard MC@NLO 
matching\footnote{\label{ftn:Gf}Interestingly, and contrary to the case 
of MC@NLO, this property would hold even if $d\sigmaMC$ and $d\sigmaNLOE$
did {\em not} have the same behaviours in such regions, which is appealing
in view of the patterns of soft emissions by MCs, and their implications
for NLO matchings~\cite{Frixione:2002ik}. We shall not elaborate on this
point any further here, and only briefly comment on it at the end of
sect.~\ref{sec:Dcostr}.}. Conversely, we know that $\clH$ events are the 
sole responsible for giving the NLO-accurate shapes and normalisations in 
hard-emission regions. Thus, we must have:
\beq
\Delta\;\longrightarrow\;1
\;\;\;\;\;\;\;\;\;\;
{\rm hard~regions}.
\label{Dcond2}
\eeq
Equations~(\ref{Dcond1}) and~(\ref{Dcond2}) are strongly reminiscent of 
the behaviour of a Sudakov form factor. Indeed, we shall later give the 
definition of $\Delta$ in terms of a combination of MC no-emission 
probabilities (which are governed by Sudakovs)\footnote{We note that 
Sudakov form factors have been employed in exclusive $\clH$ events in 
the context of a merging (as opposed to matching) procedure in 
ref.~\cite{Hoeche:2012yf}; however, no connection has been made there 
with the reduction of negative weights, and we are unable to say whether 
it actually occurs, and if so to which extent, in that method.}. Before 
going into further details, we note that Sudakovs have a well-defined 
perturbative expansion, such that: 
\beq
\Delta=1+\ord(\as)\,.
\label{Dexp}
\eeq
By using eq.~(\ref{Dexp}), a straightforward computation then shows that:
\beqn
d\sigmaHD&=&d\sigmaH+\ord\big(\as^{b+2}\big)\,,
\label{sHDeqsH}
\\
d\sigmaSD&=&d\sigmaS+\ord\big(\as^{b+2}\big)\,,
\label{sSDeqsS}
\eeqn
which imply that the generating functionals of eqs.~(\ref{genMCatNLO})
and~(\ref{genMCatNLOD}) have the same expression at the NLO (while in 
general they differ at the NNLO and beyond). Furthermore:
\beq
\sigma_{\rm NLO}\equiv
\int_{\meas_{n+1}}\left(d\sigmaS+d\sigmaH\right)=
\int_{\meas_{n+1}}\left(d\sigmaSD+d\sigmaHD\right)\,,
\label{NLOnorm}
\eeq
where $\sigma_{\rm NLO}$ is the total NLO cross section, prior to any
acceptance cuts.

We point out that eq.~(\ref{Dexp}), which obviously is {\em not} a 
property that is uniquely associated with a Sudakov, is a sufficient 
condition for eqs.~(\ref{sHDeqsH})--(\ref{NLOnorm}) to be fulfilled.
However, eq.~(\ref{Dexp}) is not sufficient for $\GenNLO$ and $\GenNLOD$ 
to produce comparable physical results\footnote{In other words, not
only results associated with a formal perturbative expansion, but also
those at the observable level that include shower effects.}; for this 
to happen other conditions, in particular that of eq.~(\ref{Dcond2})
and the requirement that \mbox{$0\le\Delta\le 1$}, are needed too. Taken 
together, then, all of these conditions constrain the functional form of
$\Delta$ to be, if not that of a Sudakov, at least Sudakov-like. Indeed, 
in order to sketch out the basic physics ideas that underpin the $\aNLOD$ 
prescription, we need only assume that the dependence of such form of $\Delta$ 
upon the kinematical $(n+1)$-body degrees of freedom can be parametrised 
in terms of two scales\footnote{In the fully realistic case we shall 
soon discuss, these two scales will be replaced by two {\em sets} 
of scales.\label{ft:sss}}, as follows:
\beq
\Delta\big(\confH\big)=\Delta\big(t(\confH),\mu^2(\confS)\big)\,.
\label{Dsimp}
\eeq
Here, $\confS$ denotes the $n$-body configuration underlying the
given $\confH$ (see eq.~(\ref{KSvsH})). We call $\mu^2(\confS)$ and
$t(\confH)$ the {\em starting} and {\em stopping} scales (squared), 
respectively. We require them to have the following properties:
\beqn
\mu(\confS)&\sim& M_H\,,
\label{start}
\\
\sqrt{t(\confH)}&\ll& M_H
\;\;\;\;\;\;\;\;\;\;
\confH~{\rm soft/collinear}\,,
\label{stop1}
\\
\sqrt{t(\confH)}&\sim& M_H
\;\;\;\;\;\;\;\;\;\;\;
\confH~{\rm intermediate/hard}\,.
\label{stop2}
\eeqn
Then, owing to the properties of eq.~(\ref{Dsimp}),
eqs.~(\ref{start}) and~(\ref{stop1}) imply eq.~(\ref{Dcond1}),
while eqs.~(\ref{start}) and~(\ref{stop2}) imply eq.~(\ref{Dcond2}).
Furthermore, by comparing eqs.~(\ref{stop1}) and~(\ref{stop2}) with
eqs.~(\ref{pull1}) and~(\ref{pull2}) one sees that the values assumed
by the stopping scale and the pull are correlated. From the classification
of negative-weight events given in sect.~\ref{sec:negw}, it then follows
that eq.~(\ref{genMCatNLOD}) is expected to reduce the number of both 
\nclo\ and \nclt\ events.

We can now return to a point raised in sect.~\ref{sec:intro},
namely that the $\ord(\as^{b+2})$ and beyond terms in an MC@NLO cross
section are of purely MC origin, i.e.~they stem from the soft and collinear
regions in the multi-parton radiation phase space.
This is guaranteed in $\aNLOD$ as well if it is the MC one matches
to that provides the ingredients\footnote{These considerations remain
valid if one replaces the ``MC'' with suitable resummed computations.
Conversely, $\Delta$ must not be defined through quantities that depend
on hard emissions (such as the exact matrix elements relevant to fixed-order
perturbative calculations).} for the construction of $\Delta$.
This means the definition of both the Sudakovs and the stopping scales;
as far as the starting scales are concerned, we point out that these
are arbitrary to a large extent, and under the control of the user. In the
current context, we shall therefore make sure that our choices are
consistent with the condition of eq.~(\ref{start}).

A key physics point in the actual definition of the $\Delta$ factor, 
which we shall soon give, is the relationship between eqs.~(\ref{xsecHD}) 
and~(\ref{Dcond1}). This tells one that the faster $\Delta$ approaches
zero, the smaller the number of $\clH$ events (and, thus, those of the
\nclo\ class). In other words, bearing in mind that we aim to construct
$\Delta$ with MC no-emission probabilities, we shall have a stronger
suppression of the short-distance contributions to $\clH$ events in
those regions where the MC is expected to have a higher probability
to emit. One may suspect that such a suppression, due to the overall
factor $\Delta$ in eq.~(\ref{xsecHD}), while leading to a smaller number
of \nclo\ events will also imply a larger number of \nclth\ events,
owing to the first term on the r.h.s.~of eq.~(\ref{xsecSD}), which is
positive, being suppressed as well. 
However, this is the case only very marginally (if at all), 
because the suppression of that first term is compensated by the presence
of the last term on the r.h.s.~of eq.~(\ref{xsecSD}). Such a compensation
is not exact, and the sum of these two terms tends to be still smaller 
than the MC-counterterm contribution without the $\Delta$ factor.
Crucially, however, the difference between these two expressions is much
smaller than the Born contribution, which dominates the $\clS$-event cross 
section. Therefore, ultimately the fractions of \nclth\ events are 
remarkably similar in MC@NLO and $\aNLOD$; if there is any difference
between the two, the latter tends to be slightly larger than the former.
As was already anticipated, both can be reduced by means of 
eq.~(\ref{genSevI}), i.e.~by folding, and this explains why the folding
is typically more effective in $\aNLOD$ than in MC@NLO. Namely, $\aNLOD$
has a much smaller fraction of negative-weight $\clH$ events w.r.t.~MC@NLO,
with only a slight increase of negative-weight $\clS$ events; while the
former are unaffected by folding, the latter are (in large part) 
eliminated by it.

\subsection{Construction of the $\Delta$ factor\label{sec:Dcostr}}
We have established the general guidelines for the definition of the
$\Delta$ factor. In order to proceed, we introduce the MC Sudakov, 
which we write as follows:
\beq
\Delta_a\left(\qt;\left\{Q_\beta^2,\ell_\beta\right\}_{\beta=1}^{\Nell(a)},
\Mdips\right)=\prod_{\beta=1}^{\Nell(a)}\Delta_a^{(*)}\left(\qt;Q_\beta^2,
\ell_\beta,\Mdips\right).
\label{suddef}
\eeq
Here, $a$ denotes the identity of the particle that potentially
undergoes a branching (i.e.~a gluon or a quark of given flavour).
The variable $\qt$, conventionally assumed to have canonical dimensions 
equal to two, is associated with the quantity in which the shower we match 
to is ordered (e.g.~a transverse momentum squared for $\pT$-ordered showers, 
and an angle weighted with an energy squared for angular-ordered showers).
In the following, we shall often refer to this quantity as the ``shower
variable''. We take into account the fact that a gluon (a quark or antiquark) 
has two (one) colour partners by setting:
\beqn
\Nell(a)&=&2\;\;\;\;\;\;\;\;\;\;a={\rm gluon}\,,
\label{Nellg}
\\
\Nell(a)&=&1\;\;\;\;\;\;\;\;\;\;a={\rm (anti)quark}\,.
\label{Nellq}
\eeqn
These partner(s) are the end (in the case of colour) and/or the beginning 
(in the case of anticolour) of the colour line(s) that begin or end 
at particle $a$; in eq.~(\ref{suddef}), we have denoted such lines 
by $\ell_\beta$. The upper value accessible to the shower variable
for emissions stemming from the $\ell_\beta$ colour connection has been
denoted by $Q_\beta^2$. Finally, by $\Mdips$ we have denoted the set of
any extra variables upon which the definition of the Sudakov may depend --
we shall give an explicit example in appendix~\ref{sec:PYesud}.

The r.h.s.~of eq.~(\ref{suddef}) is defined as follows:
\beq
\Delta_a^{(*)}\left(\qt;\Qt,\ell,\Mdips\right)=
\exp\left[-\frac{1}{\Nell(a)}\,\delta_a\left(\qt;\Qt,\ell,\Mdips\right)\right],
\label{sudstardef}
\eeq
where:
\beq
\delta_a\left(\qt;\Qt,\ell,\Mdips\right)=
\frac{\stepf\left(\Qt-\qt\right)}{N_a}\sum_b
\int_{\qt}^{\Qt}\frac{dt}{t}\frac{\as(t)}{2\pi}
\int_{\ep(t,\Mdips,\ell)}^{1-\ep(t,\Mdips,\ell)}
dz\,\hP_{ba}(z)\,.
\label{deltadef}
\eeq
Here, $\hP_{ba}(z)$ is the unsubtracted Altarelli-Parisi lowest-order kernel 
relevant to the branching of $a$ in which parton $b$ carries a fraction equal 
to $z$ of the longitudinal momentum of $a$. The r.h.s.~of eq.~(\ref{deltadef})
is summed over $b$, and the values assumed by $b$ must take into account
the relationships among the Altarelli-Parisi kernels. We find it convenient
to work with symmetric quantities; hence, when $a=q_f$, with $q_f$ a(n)
(anti)quark of flavour $f$, then \mbox{$b\in\{g,q_f\}$}, whereas if $a=g$,
then \mbox{$b\in\{g,u,\bar{u},\ldots\}$}. This implies that we need
to set:
\beq
N_a=2\,,
\label{Naeq2}
\eeq
irrespective of the identity of particle $a$.
In eq.~(\ref{deltadef}) we have denoted by $\ep$ the quantity the 
controls the phase-space boundaries of the $z$ integration; as the notation
suggests, such boundaries depend on the shower variable, on the variables 
in $\Mdips$, and on the colour line $\ell$; more details specific to the
case of \PYe~\cite{Sjostrand:2007gs} are given in appendix~\ref{sec:PYesud}. 
Finally we remark that, depending on the type of MC showers one considers, 
the $\stepf$ function on the r.h.s.~of eq.~(\ref{deltadef}) might have a 
more complicated functional dependence than that indicated (for example,
by enforcing the constraints relevant to angular-ordered showers), and 
because of this it might need to be moved under the integration signs. 
Likewise, the argument of $\as$ might be assigned by means of a different
prescription (see e.g.~refs.~\cite{Amati:1980ch,Catani:1990rr}). However, 
for the sake of the current argument there is no loss of generality in 
using eq.~(\ref{deltadef}).

We briefly comment on the physical meaning of eq.~(\ref{suddef}). When $a$ 
and $b$ are gluons, that equation and eqs.~(\ref{sudstardef})--(\ref{deltadef})
imply that each of the two $\Delta_a^{(*)}$ factors is responsible for
an evolution of effective strength equal to $C_{\sss A}/2$ -- in other
words, each colour line contributes ``half'' of the radiation 
pattern\footnote{Note that $C_{\sss A}/2$ and $C_{\sss F}$ coincide
at the leading-colour level -- it is the colour line, rather than the
particle identity, that controls the radiation.}.
As a consequence of that, the evolution from \mbox{$\max(\Qot,\Qtt)$} to 
\mbox{$\min(\Qot,\Qtt)$} is driven by branchings of effective
strength equal to $C_{\sss A}/2$ (and not $C_{\sss A}$). Conversely,
when $a$ is a quark $\Delta_a$ and $\Delta_a^{(*)}$ coincide, consistently
with the fact that in this case a single colour connection is relevant.
All of this stems from the leading-colour approximation which is (nowadays)
typically used by parton showers.

In order to employ the Sudakovs introduced above for the definition of
$\Delta$, we must first define the sets of the starting and stopping scales
(see eq.~(\ref{Dsimp}) and footnote~\ref{ft:sss}). The choices of the
former scales are motivated by the form of the MC counterterms:
\beq
d\sigmaMC=D\!\left(q(\confH),\mu_1,\mu_2\right)
\sum_c\sum_{\ell\in c}d\sigmaMC_{c\ell}\,,
\label{MCcnts}
\eeq
where by $q(\confH)$ we have denoted the square root
of the shower variable relevant to the kinematic configuration $\confH$,
for the branching in which the FKS parton $i$ and its sister $j$ emerge
from their mother\footnote{In general, $q(\confH)$ might also depend on
the end(s) of the colour line(s) to which the mother is connected. This 
dependence is trivial for all of the MCs used in \aNLOs\ at present
(among which is \PYe\ with the {\em global recoil} option).}. The reader 
can find more details on eq.~(\ref{MCcnts}) in ref.~\cite{Alwall:2014hca}
(see in particular eqs.~(2.111)--(2.116) there). Here, it suffices to
say that by \mbox{$d\sigmaMC_{c\ell}$} we have denoted the contribution
to the MC counterterms due to a given colour flow $c$ and colour 
line $\ell$; the former is identified with the set of its colour lines 
\mbox{$c=\{\ell_1,\ldots\ell_m\}$}. In turn, each of these lines is 
conveniently represented by an ordered pair of labels, with the first 
(second) being associated with the beginning (end) of the line, and the 
colour (as opposed to the anticolour) flowing from the beginning of the 
line to its end\footnote{Note that, when working with a given process and 
kinematical configuration as we are doing here, one can unambiguously choose 
a labeling convention.}. We point out that such colour flow and colour lines
are relevant to \mbox{$2\to n$} configurations. The function $D$ in 
eq.~(\ref{MCcnts}) is largely arbitrary, and is parametrised as follows:
\beqn
D(\mu;\mu_1,\mu_2)=\left\{
\begin{array}{ll}
1               &\phantom{aaaa} \mu\le\mu_1\,,\\
{\rm monotonic} &\phantom{aaaa} \mu_1<\mu\le\mu_2\,,\\
0               &\phantom{aaaa} \mu > \mu_2\,,\\
\end{array}
\right.
\label{Ddef}
\eeqn
with $\mu_\alpha$, $\alpha=1,2$, two parameters of the order of the 
hardness of the process\footnote{For example, in the current version
of \aNLOs, by default $f_1=0.1$, $f_2=1$, and $R=H_{\sss T}/2$.}:
\beq
\mu_\alpha=f_\alpha\,R\,,
\;\;\;\;\;\;\;\;
f_\alpha=\ord(1)\,,
\;\;\;\;\;\;\;\;
R\sim M_H\,.
\label{f1f2}
\eeq
The presence of the function $D$ in eq.~(\ref{MCcnts}) implies that the 
shower starting scales associated with unweighted events are generated 
by means of the following formula:
\beq
\mu=D^{-1}(r;\mu_1,\mu_2)\,,
\label{murdef}
\eeq
with $r\in [0,1]$ a flat random number.

The procedure outlined above motivates the definition of the 
starting scales for the $\aNLOD$ prescription. Firstly, we adopt
the following rule:
\begin{itemize}
\item[{\em R.2}:] Among the colour flows that contribute to eq.~(\ref{MCcnts}),
we select one of them at random by using $\sum_\ell d\sigmaMC_{c\ell}$ 
as relative probabilities, and denote it by $d$. Henceforth, by $\ell_\beta$ 
(\mbox{$1\le\beta\le m$}) we understand the colour lines that belong 
to such a flow.
\end{itemize}
Secondly, we generalise the reference scale $R$ introduced in
eq.~(\ref{f1f2}) by turning it into a set of scales, as follows:
\beq
R_{\bk\bp}^2=(-1)^{\varsigma(\bk)+\varsigma(\bp)}
\Big(\varsigma(\bk)\,\bK_{\bk}+\varsigma(\bp)\,\bK_{\bp}\Big)^2
\;\;\;\;\;\;\Longleftrightarrow\;\;\;\;\;\;
{\rm if}~~\exists\ell_\beta
\;\;\;{\rm s.t.}\;\;\;
\bk\in\ell_\beta~~{\rm and}~~\bp\in\ell_\beta\,,
\label{Rkpdef}
\eeq
where \mbox{$1\le\bk,\bp\le n$} are particle labels, and $\varsigma(\bk)=1$ 
if the particle with label $\bk$ is in the final state, while if it is
in the initial state the two choices $\varsigma(\bk)=\pm 1$ are both 
sensible; in the simulations of this paper we shall use $\varsigma(\bk)=1$,
keeping the option $\varsigma(\bk)=-1$ for future studies.
By convention, we denote Born-level quantities by barred symbols; thus, 
\mbox{$\confS=\{\bK_1,\ldots\bK_n\}$}, where $\bar{K}_{\bar{\alpha}}$ 
is the momentum of the particle labelled by $\bar{\alpha}$.
The condition in eq.~(\ref{Rkpdef}) states that the particles 
with labels $\bk$ and $\bp$ are colour connected in the colour 
flow $d$. Since each particle has either one colour partner (if it is a quark
or an antiquark) or two of them (if it is a gluon), it follows that depending 
on the process there are at least $n$ and at most $2n$ non-trivial reference 
scales defined by eq.~(\ref{Rkpdef}). We can now introduce the 
analogues of the $\mu_\alpha$ parameters of eq.~(\ref{f1f2}), namely:
\beq
\mu_{\alpha,\bk\bp}=f_\alpha\,R_{\bk\bp}\,,
\;\;\;\;\;\;\;\;\;\;
\alpha=1,2\,,
\label{f1f2mat}
\eeq
and finally the sought starting scales by analogy with 
eq.~(\ref{murdef}):
\beq
\mu_{\bk\bp}=D^{-1}(r_{\bk\bp};\mu_{1,\bk\bp},\mu_{2,\bk\bp})\,,
\label{startSdef}
\eeq
where $r_{\bk\bp}$ are flat random numbers.
As far as the stopping scales are concerned, we simply define them as the 
square roots of the MC shower variables associated with the given 
kinematical configuration. More precisely, at given $\bk$ and $\bp$, 
we denote by:
\beq
t_{\bk\bp}
\;\;\;\;\;\;\Longleftrightarrow\;\;\;\;\;\;
{\rm if}~~\exists\ell_\beta
\;\;\;{\rm s.t.}\;\;\;
\bk\in\ell_\beta~~{\rm and}~~\bp\in\ell_\beta
\label{stopSdef}
\eeq
the shower variable relevant to the emission of the FKS parton from 
the branching of parton $\bk$, when such a parton is colour-connected
to parton $\bp$. It is important to note that while for a specific
$\bk$ the branching parton coincides with the mother of the FKS parton
and its sister, in general this is not the case: $\bk$ is the mother
of the FKS parton and of another parton, not necessarily equal to
the sister of the FKS parton.
The definition in eq.~(\ref{stopSdef}) implies that there are 
as many non-trivial stopping scales as there are starting scales.
Note that in general:
\beq
t_{\bk\bp}\ne t_{\bp\bk}\,.
\label{stopSasym}
\eeq
It is useful to have a closer look at eq.~(\ref{stopSasym}) by means
of an example. Consider a Drell-Yan process, which at the Born level 
features a quark and an antiquark (labelled by $1$ and $2$, respectively) 
in the initial state, with a single colour line $(1,2)$. Suppose that a gluon 
is emitted, and employ a $\pT$-ordered MC such as \PYe. Thus, $t_{12}$
is the gluon transverse momentum squared {\em relative} to the quark, 
while $t_{21}$ is the same quantity relative to the antiquark. If for 
example the kinematical configuration is such that the gluon is collinear 
to the quark, and therefore anti-collinear to the antiquark, then 
\mbox{$t_{12}\ll t_{21}$}.

In order to arrive at the definition of no-emission probabilities
which will enter the sought $\Delta$ factor, we finally need to introduce
some auxiliary quantities. We denote by:
\beq
\ccol_\gamma(\bk)\,,\;\;\;\;\;\;\;\;
1\le\gamma\le\Nell(\ident_{\bk})\,,
\label{ccoldef}
\eeq
the labels of the particles\footnote{In the case of $\bk$ being a gluon,
the choice of which particle is the first ($\gamma=1$) or the second
($\gamma=2$) colour partner is arbitrary. None of the formulae presented
later depends on such a choice.} which are colour connected with the particle
labelled $\bk$ and, following the conventions of ref.~\cite{Alwall:2014hca}, 
by $\ident_{\bk}$ the identity of such a particle; $\Nell(\ident_{\bk})$ 
is set according to eqs.~(\ref{Nellg}) and~(\ref{Nellq}).
 One must bear in mind that we are working at a fixed colour flow ($d$) here; 
thus, the colour partners in eq.~(\ref{ccoldef}) are unambiguously defined. 
This implies that, for a given $\gamma$, there exists a single colour line 
to which both $\bk$ and $\ccol_\gamma(\bk)$ belong. We denote such a line 
as follows:
\beq
\ell\!\left[\bk,\ccol_\gamma(\bk)\right]\equiv\ell_\beta\,,
\;\;\;\;\;\;\;\;\exists!\,\ell_\beta
\;\;\;{\rm s.t.}\;\;\;
\ell_\beta=\left(\bk,\ccol_\gamma(\bk)\right)
\;\;\;{\rm or}\;\;\;
\ell_\beta=\left(\ccol_\gamma(\bk),\bk\right)\,.
\label{elldef}
\eeq
If $\bk$ labels a quark or an antiquark, we construct its no-emission
probability as follows:
\beq
\Pi_{\bk}=
F_{\bk}\!\left(t_{\bk \ccol_1\!(\bk)};\mu_{\bk \ccol_1\!(\bk)}^2\right)\,
\Delta_{\ident_{\bk}}\!\left(t_{\bk \ccol_1\!(\bk)};\mu_{\bk \ccol_1\!(\bk)}^2,
\ell\!\left[\bk,\ccol_1(\bk)\right],\Mdips\right),
\label{NEPq}
\eeq
where:
\beqn
F_{\bk}(\qt,\Qt)=
\left\{
\begin{array}{ll}
f_{\ident_{\bk}}^{(\bk)}(x,\qt)\big/f_{\ident_{\bk}}^{(\bk)}(x,\Qt) 
\,\stepf\!\left(\Qt-\qt\right)+\stepf\!\left(\qt-\Qt\right)
& \phantom{aaaa} 1\le\bk\le 2\,,\phantom{aa}\\
1  &\phantom{aaaa} \bk>2\,.\\
\end{array}
\right.
\label{Fdef}
\eeqn
In eq.~(\ref{Fdef}) $f_{\ident_{\bk}}^{(\bk)}$ is the PDF of parton 
$\ident_{\bk}$ inside the hadron coming from the left ($\bk=1$) or from 
the right ($\bk=2$). The momentum fraction $x$ that enters the PDF is the 
same as that which is used in the evaluation of the MC counterterms (see 
ref.~\cite{Alwall:2014hca} for more details). Note that, owing to
eqs.~(\ref{deltadef}) and~(\ref{Fdef}), we have:
\beq
{\rm if}~~~~t_{\bk \ccol_1\!(\bk)}\ge\mu_{\bk \ccol_1\!(\bk)}^2
\;\;\;\;\;\;\Longrightarrow\;\;\;\;\;\;
\Pi_{\bk}=1\,,
\label{Pikeqoq}
\eeq
which formalises the notion that there is no parton-shower emission in the 
hard region, whose lower boundary coincides by definition with the starting 
scale.

The case where $\bk$ labels a gluon is more complicated. In a non-matched
case, the MC uses the Sudakov of eq.~(\ref{suddef}) (or rather, a
no-emission probability that differs from the Sudakov in a way which
is irrelevant in this discussion) to extract randomly a value $\qt$ of
the shower variable. In doing so, the two colour lines bid to emit.
Once $\qt$ has been generated, the corresponding kinematical configuration
is reconstructed; however, in order to do so the MC must choose a definite
colour line, since in general the kinematic configurations associated
with the same $\qt$ that emerge from different colour lines are different.
In the present case, this logic must be reversed, since we start from a
definite kinematical configuration; this implies that the two colour lines
might give rise to different values of the shower variable. We stress that
this need not be so; but we work in this scenario in order to be as 
general as possible. We then define the analogue of the no-emission
probability of eq.~(\ref{NEPq}) in the case of a gluon as follows:
\beqn
\Pi_{\bk}&=&\frac{g_1}{g_1+g_2}
F_{\bk}\!\left(t_{\bk \ccol_1\!(\bk)};\mu_{\bk \ccol_1\!(\bk)}^2\right)\,
\Delta_{\ident_{\bk}}\!\left(t_{\bk \ccol_1\!(\bk)};
\left\{\mu_{\bk \ccol_\gamma\!(\bk)}^2,
\ell\!\left[\bk,\ccol_\gamma(\bk)\right]\right\}_{\gamma=1}^{2},
\Mdips\right)
\nonumber\\*&+&
\frac{g_2}{g_1+g_2}
F_{\bk}\!\left(t_{\bk \ccol_2\!(\bk)};\mu_{\bk \ccol_2\!(\bk)}^2\right)\,
\Delta_{\ident_{\bk}}\!\left(t_{\bk \ccol_2\!(\bk)};
\left\{\mu_{\bk \ccol_\gamma\!(\bk)}^2,
\ell\!\left[\bk,\ccol_\gamma(\bk)\right]\right\}_{\gamma=1}^{2},
\Mdips\right),\phantom{aa}
\label{NEPg}
\eeqn
where:
\beq
g_\gamma=\delta_{\ident_{\bk}}\!\left(t_{\bk \ccol_\gamma\!(\bk)};
\mu_{\bk \ccol_\gamma\!(\bk)}^2,
\ell\!\left[\bk,\ccol_\gamma(\bk)\right],\Mdips\right),
\eeq
and $\delta_\ident$ is defined in eq.~(\ref{deltadef}). Apart from the 
$g_\gamma$-dependent prefactor, each of the two terms on the r.h.s.~of
eq.~(\ref{NEPg}) is the standard gluon no-emission probability. The
difference between them is solely due to the choice of the stopping
scale that enters the PDF factor and the Sudakov, $t_{\bk \ccol_1\!(\bk)}$
versus $t_{\bk \ccol_2\!(\bk)}$, and to that of the starting scale that enters 
the PDF, $\mu_{\bk \ccol_1\!(\bk)}^2$ versus $\mu_{\bk \ccol_2\!(\bk)}^2$.
The $g_\gamma$-dependent prefactors have the role of 
combining these two no-emission probabilities by means of a
weighted average, in which the weights are MC-driven, and are such
that the term associated with the smallest stopping scale is multiplied by 
the largest factor. The analogue for eq.~(\ref{NEPg}) of eq.~(\ref{Pikeqoq})
reads as follows:
\beq
{\rm if}~~~~\min\left(t_{\bk \ccol_1\!(\bk)},t_{\bk \ccol_2\!(\bk)}\right)\ge
\max\left(\mu_{\bk \ccol_1\!(\bk)}^2,\mu_{\bk \ccol_2\!(\bk)}^2\right)
\;\;\;\;\;\;\Longrightarrow\;\;\;\;\;\;
\Pi_{\bk}=1\,.
\label{Pikeqog}
\eeq
We must also mention the fact that it is possible that, for given
kinematical configuration, emitter, and colour partner (or partners
in the case of a gluon), the FKS parton is in an MC dead zone (i.e.~a 
phase-space region which is not accessible by MC radiation; when $\bk$ 
labels a gluon, this condition is fulfilled only if the FKS parton is
in a dead zone according to {\em both} of the colour connections of $\bk$). 
In such a case we set $\Pi_{\bk}=1$, since that kinematical configuration 
could not have been generated by the MC.

By using eqs.~(\ref{NEPq}) and~(\ref{NEPg}) we can finally define
the $\Delta$ factor we need for the $\aNLOD$ matching:
\beq
\Delta=\prod_{\bk=1}^n\Pi_{\bk}\,.
\label{Deltafin}
\eeq
Although the construction of the no-emission probabilities that
enter eq.~(\ref{Deltafin}) is involved, the physics content of the
$\Delta$ factor is easy to understand, because it is quite literally
that of eq.~(\ref{Dsimp}). In fact, in a typical situation all but
one of the $\Pi_{\bk}$ terms on the r.h.s.~of eq.~(\ref{Deltafin})
are of order one. Furthermore, the single no-emission probability
for which $\Pi_{\bk}\ll 1$ is most likely that where $\bk$ identifies
the particle that branches into the FKS parton and its sister. This
is because by working in the FKS sector defined by a given $\Sfun_{ij}$
function, one suppresses all collinear configurations except that
where \mbox{$K_i\!\parallel\!K_j$}~\cite{Frixione:1995ms,
Frixione:1997np}; for the former the stopping scales are 
``large'' (i.e.~as in eq.~(\ref{stop2})), while for the latter
they are ``small'' (i.e.~as in eq.~(\ref{stop1})). It is interesting 
to observe that in MC@NLO there is a non-zero probability that, 
within the $\Sfun_{ij}$ sector, the FKS parton is closer to a particle 
different from its sister (we denote by $p$ its label) than to its 
sister; let us call anticollinear such a configuration. 
This behaviour is driven by the fact that while the real-emission 
component of the short-distance cross sections is multiplied by 
$\Sfun_{ij}$, the MC counterterms are not since, at variance with
the matrix elements, in anticollinear configurations they are naturally
suppressed. However, they are not equal to zero, which induces the small
but non-zero probability, alluded to before, of generating events there.
Furthermore, since these are $\clH$ events predominantly stemming from
the MC counterterms, it is likely that they will have negative weights
(see eq.~(\ref{xsecH})). While this is the mechanism at work in the
context of MC@NLO, in $\aNLOD$ these anticollinear configurations are 
suppressed -- by the no-emission probability $\Pi_{\bk}$ with $\bk$ 
now identifying the particle that branches into the FKS parton
and the particle labelled by $p$. Therefore, thanks to the symmetry of 
eq.~(\ref{Deltafin}), the $\aNLOD$ prescription reduces the number of 
events of class \nclo\ in both collinear and anticollinear configurations.
This is just as well, because the separation into collinear and anticollinear
regions stems from adopting the FKS subtraction for writing the 
short-distance cross sections, whereas the ideas that underpin
$\aNLOD$ are independent of the FKS formalism. 

The $\Delta$ factor of eq.~(\ref{Deltafin}) may feature a suppression due 
to multiple no-emission probabilities in the case of soft configurations.
We expect the strength of this suppression to depend significantly on
the MC one matches to. For example, soft emissions in angular-ordered MCs
have {\em typically} a pattern whereby the contributions due to different 
emitters have a smaller overlap than in the case of $\pT$-ordered showers;
this implies that the value of $\Delta$ associated with soft configurations
will be typically larger (i.e.~resulting in a smaller suppression) in 
angular-ordered showers than in $\pT$-ordered ones. In any case, the
suppression induced by $\Delta$ is interesting because of the technical
issues related to soft events in MC@NLO (see in particular appendix~A.5 of 
ref.~\cite{Frixione:2002ik} and eq.~(2.117) of ref.~\cite{Alwall:2014hca}).
By construction, these issues become much less relevant in the context
of $\aNLOD$, to the extent that in such a formulation one might get rid
of the function ${\cal G}$ in the definition of the MC counterterms, 
since $\Delta$ vanishes in the regions where ${\cal G}$ is non-trivial -- 
it is the kinematics, and not the colour structure, that dictates where 
a cross section diverges; see also footnote~\ref{ftn:Gf}.

We conclude this discussion with a comment on the colour structure of the 
$\aNLOD$ cross section. As is the case for all quantities stemming from
a random unweighting of competing contributions, the $\Delta$ factor
constructed with $d$ will also multiply short-distance terms that feature
colour flows different from $d$. This is perfectly fine, and in fact
every MC is based on several applications of a similar nature. In the
specific example of $\Delta$, however, a possible alternative procedure
that would allow one to bypass the unweighting would be that of decomposing
{\em all} matrix elements as is done for Born-level quantities in 
standard MCs, e.g.~by applying the prescription of ref.~\cite{Odagiri:1998ep}.
We shall not pursue any further this option in the present paper.

\subsection{Assignments of scales in hard events\label{sec:LHEscales}}
The definition of the $\Delta$ factor in the $\aNLOD$ prescription naturally 
suggests how to improve the structure of the hard scales which are associated
with hard events. At present, in NLO simulations a single hard scale
per event is employed; we propose to exploit the starting and stopping
scales in order to assign either one (for quarks) or two (for gluons)
hard scale(s) {\em per particle}. In particular, in $\clS$ events the 
scales associated with the particle labelled by $\bk$ are the following:
\beq
\bk~{\rm particle}\,,\;\;\;\clS~{\rm events}
\;\;\;\;\;\;\longrightarrow\;\;\;\;\;\;
\left\{\mu_{\bk \ccol_\gamma\!(\bk)}\right\}_{\gamma=1}^{\Nell(\ident_{\bk})}\,.
\label{Sevsc}
\eeq
In other words, in the case of $\clS$ events each particle is associated
with as many hard scales as its colour connections; each of these scales
is equal to the starting scale stemming from the corresponding colour line.

In the case of $\clH$ events, the obvious analogue of eq.~(\ref{Sevsc})
is that in which one sets the hard scales equal to the stopping scales.
However, in order to do so some care is required: in fact, although
stopping scales are constructed in terms of \mbox{$(n+1)$}-body kinematical
configurations, the particle indices they depend upon are those of an
$n$-body process (see eq.~(\ref{stopSdef})). 
We start by observing that, within a given FKS
sector (i.e.~for a given $\Sfun_{ij}$ function) the particle indices
at the \mbox{$(n+1)$}-body and $n$-body levels can unambiguously be mapped
onto each other: we simply use the same symbol, barred (unbarred) in the 
case of the $n$-body (\mbox{$(n+1)$}-body) process to denote corresponding
particles (for example, $\bk$ and $k$ denote the index of a particle
before and after, respectively, the branching that turns an $n$-body 
configuration into an \mbox{$(n+1)$}-body one). The FKS parton $i$ and
its sister $j$ constitute a special case, since they have no analogues
at the $n$-body level; there, their mother appears instead. We denote 
it by $\mother$.

The colour flow of each $\clH$ event is constructed from that
of the underlying $n$-body configuration ($d$), by breaking the colour 
line (in the case of a quark), or one of the two colour lines (in the case 
of a gluon; for a $g\to q\bar{q}$ branching, the lines ``split''), 
associated with the branching of $\mother$;
the other colour lines that belong to $d$ are left untouched (bar the
obvious particle relabeling). This operation is fully deterministic,
except in the case of a $g\to gg$ branching, for which two different
$\clH$ colour flows can be constructed: we choose one of them randomly
(with equal probabilities).

After the definition of the $\clH$-event colour flow, in keeping with
eq.~(\ref{Sevsc}) we assign one or two hard scale(s) per particle,
according to the colour connection(s) of such a particle. We do that by
a suitable mapping of the $n$-body indices relevant to the stopping scales
onto the \mbox{$(n+1)$}-body ones. Explicitly:
\beqn
&&{\rm if}~~\exists\,\ell\!\left[k,p\right]
\;\;\;\;\;\Longrightarrow\;\;\;\;\;
t_{kp}=t_{\bk\bp}\,\,
\;\;\;\;\;\;\;\;\phantom{a}\;
k\ne i\,,k\ne j\,,p\ne i\,,p\ne j\,,
\label{Ht1}
\\
&&{\rm if}~~\exists\,\ell\!\left[k,j\right]
\;\;\;\;\;\Longrightarrow\;\;\;\;\;
t_{kj}=t_{\bk\mother}\,\,
\;\;\;\;\;\;\;\;
k\ne i\,,k\ne j\,,
\label{Ht2}
\\
&&{\rm if}~~\exists\,\ell\!\left[k,i\right]
\;\;\;\;\;\Longrightarrow\;\;\;\;\;
t_{ki}=t_{\bk\mother}\,\,
\;\;\;\;\;\;\;\;
k\ne i\,,k\ne j\,,
\label{Ht3}
\\
&&{\rm if}~~\exists\,\ell\!\left[i,k\right]
\;\;\;\;\;\Longrightarrow\;\;\;\;\;
t_{ik}=t_{\mother\bp}\,\,
\;\;\;\;\;\;\;\;
\bp~~{\rm s.t.}~~i~{\rm colour~connected~to}~p\,,
\label{Ht4}
\\
&&{\rm if}~~\exists\,\ell\!\left[j,k\right]
\;\;\;\;\;\Longrightarrow\;\;\;\;\;
t_{jk}=t_{\mother\bp}\,\,
\;\;\;\;\;\;\;\;
\bp~~{\rm s.t.}~~i~{\rm colour~connected~to}~p\,.
\label{Ht5}
\eeqn
In eqs.~(\ref{Ht1})--(\ref{Ht5}) we have employed the same symbol
$\ell\!\left[a,b\right]$ introduced in eq.~(\ref{elldef}) to denote
the colour line (which, in this case, belongs to the $\clH$-event
colour flow) that connects particle $a$ and particle $b$, so that
$t_{ab}$ is a meaningful quantity. Owing to the construction of the 
$\clH$-event colour flow described before, it is easy to see that 
the pairs of indices on the r.h.s.'s of eqs.~(\ref{Ht1})--(\ref{Ht5}) 
are also associated with two particles which are colour connected at 
the $n$-body level; therefore, the corresponding scales belong to 
the set of stopping scales previously determined. We point out that
the conditions in eqs.~(\ref{Ht2}) and~(\ref{Ht3}), and those in
eqs.~(\ref{Ht4}) and~(\ref{Ht5}), need not be simultaneously satisfied;
in particular, this is never the case if $k$ labels a quark. When 
they are, the corresponding scales are assigned the same value, in
turn equal to a stopping scale relevant to the mother of the FKS
parton and its sister. This is in keeping with the special roles
played by the latter partons, a fact that has been already exploited 
in the construction of the $\clH$-event colour flow.

In summary, in $\clH$ events we set the scales associated with the 
particle labelled by $k$ as follows:
\beq
k~{\rm particle}\,,\;\;\;\clH~{\rm events}
\;\;\;\;\;\;\longrightarrow\;\;\;\;\;\;
\left\{\sqrt{t_{k\ccol_\gamma\!(k)}}\right\}_{\gamma=1}^{\Nell(\ident_k)}\,,
\label{Hevsc}
\eeq
where $\ccol_\gamma\!(k)$ now denotes the label of the $\gamma^{th}$ 
colour partner of the particle with label $k$ according to the
$\clH$ colour flow.

The scale structure defined by eqs.~(\ref{Sevsc}) and~(\ref{Hevsc})
can be easily included in Les Houches event (LHE henceforth) 
files~\cite{Alwall:2006yp,Butterworth:2010ym}, which in turn are
fully compatible with modern MCs; more details on this point are
given in sect.~\ref{sec:impl}. This is expected to give the MC
options for showering scales which are, on an {\em event-by-event basis},
more sensible than those relevant to single-scale hard events
(i.e.~those which are presently in use). The reasons that inform this 
argument are general, and thus valid for both $\clS$ and $\clH$ events, 
but are easier to understand if one uses an $\clH$ event as a practical
example. This is because particles in $\clS$ events tend to be hard
and well separated from each other, while this is typically not the
case for $\clH$ events -- in other words, while both $\clS$ and $\clH$
events constitute multi-scale environments, the hierarchy among the
scales of the latter events will typically be stronger than that of 
the former ones. Let us then consider an $\clH$-event configuration
in which two particles have a small angular separation. 
By adopting an MC-driven picture,
these two particles are naturally seen as emerging from the branching
of a mother particle, and thus any subsequent shower emission off them 
should use a ``small'' scale (owing to the small angular separation) as 
a reference. Conversely, shower emissions from the other particles of the 
event (at least those not colour-connected with the former two) would tend
to use a ``large'' scale (these other particles being well-separated from
each other and hard) as a reference. This small-scale large-scale hierarchy
may be handled in single-scale LHE files in a variety of ways (by means of 
an average assignment, by using a biased random assignment, and so forth),
but it is unavoidable that, for some individual events and subsequent
shower histories, a sub-optimal choice will be made, with a better 
physical modeling recovered, in principle, only with large statistics.
This issue is simply not relevant if LHE files incorporate a multi-scale
structure, such as the one we advocate in eqs.~(\ref{Sevsc}) and~(\ref{Hevsc})
and implement in $\aNLOD$.

\section{Implementation\label{sec:impl}}
In order to achieve an $\aNLOD$ matching, any existing MC@NLO program
must be given the capability to compute the $\Delta$ factor of
eq.~(\ref{Deltafin}). As is discussed in sect.~\ref{sec:mcnloD},
although perturbatively there is a significant freedom in the definition 
of $\Delta$, the ideal situation is that in which it is the MC one matches 
to that provides the ingredients that enter $\Delta$. From the conceptual
point of view, this is quite analogous to the situation of the
MC counterterms, which are constructed {\em analytically} through
a formal expansion in $\as$ of the MC cross sections. However, although some
of the quantities that help define $\Delta$ also enter the MC counterterms
(specifically, the stopping scales and dead zones associated with $\bk$
being the label of the mother of the FKS parton and its sister), the
majority of them would have to be computed from scratch if one had to
follow the same strategy that relies on analytical results. We believe
this to be an error-prone procedure that also lacks flexibility, and
we therefore pursue a different approach (which, in due time, could
also be applied the MC counterterms themselves). Namely, we construct
$\Delta$ {\em numerically}, by essentially employing the same modules
as the MC does when showering.
 
We point out that this approach entails non-trivial modifications
to the structure of the interface between \aNLOs\ and the MC. In particular,
in the current MC@NLO implementation of \aNLOs\ (i.e.~one based on the 
analytical forms of the MC counterterms), the phase-space integration of the
short-distance cross sections and the unweighting of the events (thus,
the creation of the LHE files) do not require the use of the MC. The MC
is run independently of \aNLOs\ and after it, using as inputs the LHE
files produced by the latter program. Conversely, with the $\aNLOD$ 
implementation envisaged previously, MC routines must be called
for each phase-space point generated by the integration routine.

In view of this, the \aNLOs\ and \PYe\ versions used to obtain the 
results presented here include the following features:
\begin{itemize}
\item The executable relevant to the phase-space integration and
unweighting-event phases is constructed by linking the \aNLOs\
and \PYe\ programs.
\item New modules have been included in \PYe, that call native \PYe\
modules relevant to showering, and which in turn are called by \aNLOs\ 
for the construction of $\Delta$ in the definition of the short-distance
cross sections $d\sigmaHD$ and $d\sigmaSD$. For any given phase-space point,
these new modules return the stopping scales and the information on the
dead zones.
\item Analogous modules are constructed that return the \PYe\ Sudakovs.
However, in view of the fact that the Sudakovs depend on kinematical 
configurations in a way that can be parametrised, such modules are
called prior to the phase-space integration, by a program that defines
the Sudakovs as look-up tables. It is such tables that are used by
\aNLOs\ during integration and unweighting, which allows one to save
a significant amount of CPU time. More details on this procedure,
that we dub {\em pre-tabulation}, are given in appendix~\ref{sec:PYesud}.
\end{itemize}
As was already said, this structure paves the way to the numerical
construction of the MC counterterms themselves. We have not pursued 
this option in the course of the present work, where we employ the same 
MC counterterms as those originally computed for the MC@NLO matching.

For the MC@NLO and $\aNLOD$ matchings, we have also implemented 
the $\clS$-event generating functional according to eq.~(\ref{genSevI})
(with \mbox{$d\sigmaS\to d\sigmaSD$} there in the case of $\aNLOD$).
This is done by expressing the integral on the r.h.s.~of that equation
through Riemann sums:
\beq
\GenMC\!\left(\confS\right)\int_{\meas_r}\!d\sigmaS\simeq
\GenMC\!\left(\confS\right)
\sum_{i_\xi=1}^{n_\xi}\sum_{i_y=1}^{n_y}\sum_{i_\varphi=1}^{n_\varphi}
\frac{w_{i_\xi i_y i_\varphi}}{n_\xi n_y n_\varphi}\,
d\sigmaS\left(\confS,\xi_{i_\xi},y_{i_y},\varphi_{i_\varphi}\right).
\label{fold}
\eeq
This implies that, for each $n$-body configuration, 
\mbox{$n_\xi\times n_y\times n_\varphi$} points are generated that
span the $\meas_r$ subspace. The capability of choosing such points
(\mbox{$(\xi_{i_\xi},y_{i_y},\varphi_{i_\varphi})$}) and the
corresponding weights ($w_{i_\xi i_y i_\varphi}$) is fortunately
already available in the integration routine
({\sc\small MINT}~\cite{Nason:2007vt}) that has been adopted
in \aNLOs. Following {\sc\small MINT}, we call folding the procedure 
on the r.h.s.~of eq.~(\ref{fold}). The integers $n_\xi$, $n_y$, and
$n_\varphi$ are called folding parameters, and are under the user's
control. We shall discuss their use in sect.~\ref{sec:res}; for the
time being, note that by choosing all of the folding parameters equal
to one we recover the standard $\clS$-event generating functional.

We conclude this section with a few words on the shower scales
associated with hard events in the context of $\aNLOD$. These are
defined as is described in sect.~\ref{sec:LHEscales}, and included
in LHE files by means of the {\tt <scales>} tag. In particular, a
typical LHE will read as follows:

\vskip 0.4truecm
\hskip -0.1truecm
\begin{minipage}{0.7\textwidth}
{\tt
\begin{verbatim}
<event>
  ....
  <scales muf='1.0E+01' mur='1.0E+01' ... scalup_a_b='X' ...>
  </scales>
</event>
\end{verbatim}
}
\end{minipage}

\vskip 0.4truecm
\noindent
For each event, there are as many {\tt scalup\_a\_b} entries as is necessary,
with {\tt a} and {\tt b} being equal to either $\bk$ and $\ccol_\gamma\!(\bk)$
for $\clS$ events (see eq.~(\ref{Sevsc})), or $k$ and $\ccol_\gamma\!(k)$
for $\clH$ events (see eq.~(\ref{Hevsc}))\footnote{In other words,
\mbox{\{{\tt a},{\tt b}\}}$=$\mbox{\{{\tt I},{\tt J}\}}, where
{\tt I} and {\tt J} label two particles in the LHE file such that
\mbox{{\tt ICOLUP(K,I)}$=${\tt ICOLUP(L,J)}}, with 
\mbox{{\tt K}, {\tt L} $\in\{1,2\}$}.}. The values {\tt X} 
written above represent either the starting scales 
$\mu_{\bk \ccol_\gamma\!(\bk)}$
for $\clS$ events or the stopping scales $\sqrt{t_{k\ccol_\gamma\!(k)}}$
for $\clH$ events. As is shown in the example above, the {\tt <scales>}
tag is the last entry of a LHE. This structure is compatible with the
guidelines of the Les Houches accord~\cite{Alwall:2006yp,Butterworth:2010ym},
and with any recent \PYe\ 8.2 version.

\section{Results\label{sec:res}}
In this section we present $\aNLOD$ predictions for several hadroproduction
processes, and compare them with their MC@NLO counterparts. All of these
results have been obtained by means of \aNLOs~\cite{Alwall:2014hca} 
and \PYe~\cite{Sjostrand:2007gs}; the implementation of the $\aNLOD$
prescription in the former, and the corresponding utility modules in
the latter, will become publicly available shortly after the release 
of this paper.

We have performed our runs by adopting the default \aNLOs\ parameters;
all results are relevant to $pp$ collisions at $\sqrt{S}=13$~TeV.
The particle masses and widths are set as follows:
\beqn
&&
m_t=173~\gev\,,\;\;\;\;\;\;\phantom{aaaa}\;\,
m_H=125~\gev\,,\;\;\;\;\;\;
\\*&&
m_W=80.385~\gev\,,\;\;\;\;\;\;\phantom{a}\,
m_Z=91.188~\gev\,,
\\*&&
\Gamma_W=2.047600~\gev\,,\;\;\;\;\;\;
\Gamma_Z=2.441404~\gev\,.
\eeqn
Both the Higgs and the top-quark widths have been set equal to zero,
the latter choice being allowed by the specific production processes
we have considered, that do not feature internal top-quark propagators
which might go on-shell. We have adopted the central NNPDF2.3 PDF 
set~\cite{Ball:2012cx}, that is associated with the value
\beq
\as(m_Z)=0.119\,.
\eeq
We have also set:
\beq
G_{\sss F} = 1.16639\cdot 10^{-5}~\gev^{-2}\,,\;\;\;\;\;\;\;\;
\frac{1}{\aem}=132.507\,.
\label{aemGmu}
\eeq
The central values of the renormalisation and factorisation 
scales have been taken equal to the reference scale:
\beq
\mu=\frac{\Ht}{2}\equiv 
\frac{1}{2}\sum_i\sqrt{m_i^2+\pT^2(i)}\,,
\label{scref}
\eeq
where the sum runs over all final-state particles. In this paper, we
have not considered the theoretical systematics associated with the
variations of these scales. The \PYe\ parameters are the default ones,
with the possible exception of those settings specific to MC@NLO 
matching\footnote{See 
\href{http://amcatnlo.web.cern.ch/amcatnlo/list_detailed2.htm\#showersettings}
{http://amcatnlo.web.cern.ch/amcatnlo/list\_detailed2.htm\#showersettings}.};
these apply to $\aNLOD$ as well. The hadronisation, underlying events,
and QED showers have been turned off; the top, Higgs, and electroweak
vector bosons emerging from the hard processes have been treated
as stable, and thus left undecayed.

We have considered the following processes:
\beqn
pp &\longrightarrow& e^+ e^-\,,
\label{procee}
\\
pp &\longrightarrow& e^+ \nu_e\,,
\label{procenu}
\\
pp &\longrightarrow& H\,,
\label{procH}
\\
pp &\longrightarrow& H b \bar{b}\,,
\label{procHbb}
\\
pp &\longrightarrow& W^+ j\,,
\label{procWj}
\\
pp &\longrightarrow& W^+ t\bar{t}\,, 
\label{procWtt}
\\
pp &\longrightarrow& t\bar{t}\,, 
\label{proctt}
\eeqn
that constitute a sufficiently diverse set as far as underlying partonic
production mechanisms, complexity of phase-space, induced parton-shower 
dynamics, and behaviour under matching (including the fractions of 
negative-weight events) are concerned. This allows one to obtain a
reasonably complete comparison between MC@NLO and $\aNLOD$ results, 
as well as to have a first idea of the main features of the latter
matching prescription. 
\begin{table}
\begin{center}
\renewcommand\arraystretch{1.3}
\resizebox{\textwidth}{!}{
\begin{tabular}{l@{\qquad}r@{\% }lr@{\% }lr@{\% }l@{\qquad\quad}r@{\% }lr@{\% }lr@{\% }l}
\toprule
& \multicolumn{6}{c}{MC@NLO\qquad\qquad\qquad} &\multicolumn{6}{c}{$\aNLOD$\qquad\qquad} \\
& \multicolumn{2}{c}{111} & \multicolumn{2}{c}{221} & \multicolumn{2}{c}{441\qquad\qquad} & \multicolumn{2}{c}{$\Delta$-111} & \multicolumn{2}{c}{$\Delta$-221} & \multicolumn{2}{c}{$\Delta$-441}\\
\midrule
$pp \to e^+ e^-$ & 6.9& (1.3) & 3.5& (1.2) & 3.2& (1.1) & 5.7& (1.3) &
2.4& (1.1) & 2.0& (1.1) \\
$pp \to e^+ \nu_e$ & 7.2& (1.4) & 3.8& (1.2) & 3.4& (1.2) & 5.9& (1.3) &
2.5& (1.1) & 2.3& (1.1) \\
$pp \to H$ & 10.4& (1.6) & 4.9& (1.2) & 3.4& (1.2) & 7.5& (1.4) & 2.0&
(1.1) & 0.5& (1.0) \\
$pp \to H b \bar{b}$ & 40.3& (27) & 38.4& (19) & 38.0& (17) & 36.6& (14) &
32.6& (8.2) & 31.3& (7.2) \\
$pp \to W^+ j$ & 21.7& (3.1) & 16.5& (2.2) & 15.7& (2.1) &
14.2& (2.0) & 7.9& (1.4) & 7.4& (1.4) \\
$pp \to W^+ t\bar{t}$ & 16.2& (2.2) & 15.2& (2.1) & 15.1& (2.1) & 13.2&
(1.8) & 11.9& (1.7) & 11.5& (1.7) \\
$pp \to t\bar{t}$ & 23.0& (3.4) & 20.2& (2.8) & 19.6& (2.7) & 13.6& (1.9)
& 9.3& (1.5) & 7.7& (1.4) \\
\bottomrule
\end{tabular}
}
\caption{
\label{tab:procs}
Fractions of negative-weight events, $f$, and the corresponding 
relative costs, $c(f)$ (in round brackets), for the processes in 
eqs.~(\ref{procee})--(\ref{proctt}), computed with MC@NLO (columns 2--4) 
and with $\aNLOD$ (columns 5--7), for three different choices of the 
folding parameters.
}
\end{center}
\end{table}
The process in eq.~(\ref{procHbb}) has been computed, with 
\mbox{$m_b=4.7~\gev$}, in a four-flavour scheme; thus, there is
a slight inconsistency due to the usage of the (five-flavour scheme)
NNPDF2.3 PDFs, which is however irrelevant for the purpose of the present
study. The results of the process in eq.~(\ref{procWj}) have been obtained 
by imposing a \mbox{$\pT\ge 50~\gev$} cut on the hardest jet of the 
event; jets are reconstructed by means of 
{\sc\small FastJet}~\cite{Cacciari:2011ma},
with an $R=0.5$ anti-$\kt$ algorithm~\cite{Cacciari:2008gp}. 
We remind the reader that the starting scales are determined as is explained
in sect.~\ref{sec:Dcostr}; in particular, see eq.~(\ref{f1f2}) (for MC@NLO)
and eq.~(\ref{f1f2mat}) (for $\aNLOD$), where $f_\alpha$ are free parameters, 
whose values we are soon going to specify. In order to do that, in view of 
what is implemented in the \aNLOs\ code it is customary to define the 
$f_\alpha$'s in a redundant way, namely:
\beq
f_\alpha=\kappa\hat{f}_\alpha\,.
\label{kffdef}
\eeq
The default choices of these parameters for all of the processes in
eqs.~(\ref{procee})--(\ref{proctt}), except for that in eq.~(\ref{procHbb}),
are the following:
\beq
\kappa=1\,,\;\;\;\;\;\;
\hat{f}_1=0.1\,,\;\;\;\;\;\;
\hat{f}_2=1\,,
\label{kffval}
\eeq
while in the case of eq.~(\ref{procHbb}) we set:
\beq
\kappa=\half\,,\;\;\;\;\;\;
\hat{f}_1=0.1\,,\;\;\;\;\;\;
\hat{f}_2=1\,.
\label{kffHbb}
\eeq
The reduced value of the $\kappa$ parameter in eq.~(\ref{kffHbb}) 
w.r.t.~that of eq.~(\ref{kffval}) is in keeping with the findings
of ref.~\cite{Wiesemann:2014ioa}. In the case of MC@NLO, we use
(see eq.~(\ref{f1f2})):
\beq
R=\frac{\Ht}{2}\,,
\label{Rdef}
\eeq
which has been the \aNLOs\ default since version 2.5.3\footnote{This renders
a direct comparison of the present MC@NLO results for $Hb\bb$ production
with those of ref.~\cite{Wiesemann:2014ioa} impossible, in view of a
different choice of the scale $R$ made in that paper. However, the
combination of eqs.~(\ref{kffHbb}) and~(\ref{Rdef}) leads to predictions
which are rather similar to those obtained with the $\alpha=1/4$ choice of 
ref.~\cite{Wiesemann:2014ioa}, i.e.~the default there. See that paper for 
more details.}. 

We start by reporting, in table~\ref{tab:procs}, the overall fractions
$f$ of negative-weight events, expressed in percentage, for the processes 
in eqs.~(\ref{procee})--(\ref{proctt}). For each such fraction, we also
give the value of the corresponding relative cost, defined in 
eq.~(\ref{costdef}) and computed with $C_\pm=0$ -- these are the entries 
in round brackets. For each of the processes that we have considered, there
are six results; those in columns 2 to 4 are obtained with MC@NLO, while
those in columns 5 to 7 are obtained with $\aNLOD$. The three results
relevant to a given matching prescription (either MC@NLO or $\aNLOD$)
differ by the choices of the folding parameters
\beq
n_\xi\,,\;\;\;\;\;\;
n_y\,,\;\;\;\;\;\;
n_\varphi\,,
\label{foldpar}
\eeq
introduced in eq.~(\ref{fold}); the labellings of the columns correspond
to the three integers in eq.~(\ref{foldpar}), in that order. 
Note that \mbox{$(n_\xi,n_y,n_\varphi)=(1,1,1)$} is equivalent to not
performing any folding.

While for any given matching prescription and folding choice there
are large differences among the results, presented in table~\ref{tab:procs},
associated with the various processes considered, the pattern of the reduction 
of the fraction of negative-weight events is remarkably similar across
processes. Namely, by fixing the folding parameters $\aNLOD$ has a smaller
fraction of negative weights w.r.t.~MC@NLO; and by fixing the matching 
prescription, an increase of the folding parameters leads to a decrease
of the fraction of negative weights. This implies that, as far as the
relative cost is concerned, MC@NLO without folding and $\aNLOD$ with
maximal folding sit at the opposite ends of the spectrum, the former
being the worst (largest $c(f)$) and the latter being the best 
(smallest $c(f)$). A closer inspection of such a pattern, that involves
considering the fractions of negative weights separately for $\clS$ and
$\clH$ events (not shown in the table), confirms that the mechanism that 
underpins the reduction of negative weights is the one sketched out just 
before sect.~\ref{sec:Dcostr}; more details on this point will be presented 
in sect.~\ref{sec:ttb}, using $t\bt$ production as an example.

We point out that the qualification ``maximal'' applied above to folding
obviously refers to the set of parameter choices we have employed.
While this is far from being complete, we can confidently make a few
observations. Firstly, we have verified that by increasing
$n_\varphi$ there is a negligible reduction of negative weights, 
while of course the running time increases; this is the reason why
we have only presented $n_\varphi=1$ results. Secondly, heuristically
it appears that symmetric choices $n_\xi=n_y$ lead to smaller relative
costs than asymmetric ones $n_\xi\ne n_y$. Thirdly, while we did consider
$n_\xi=n_y>4$ for the simplest processes, the amount of reduction of
negative weights did not seem worth it, in view of the corresponding 
increase of the running time. This is likely due to the fact that, for
the simplest processes, the fraction of negative-weight events is already
quite small for our maximal-folding choice, so that any further reduction 
is difficult to achieve. Conversely, depending on the user's computing 
resources, in the case of more involved processes the exploration of 
stronger foldings than those considered here might be envisaged.

\begin{figure}[thb]
  \begin{center}
  \includegraphics[width=0.47\textwidth]{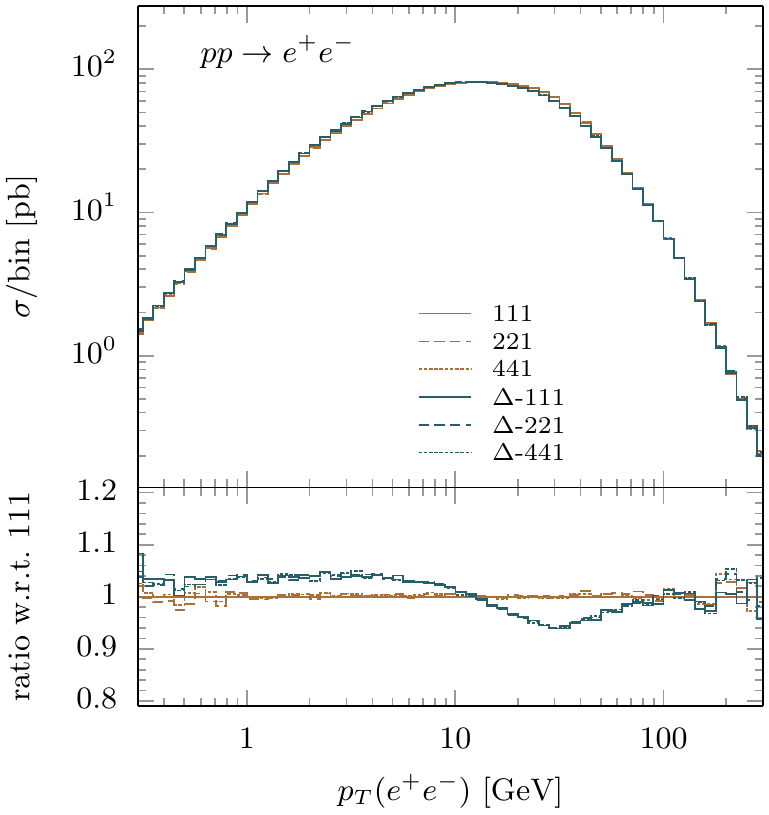}
$\phantom{a}$
  \includegraphics[width=0.47\textwidth]{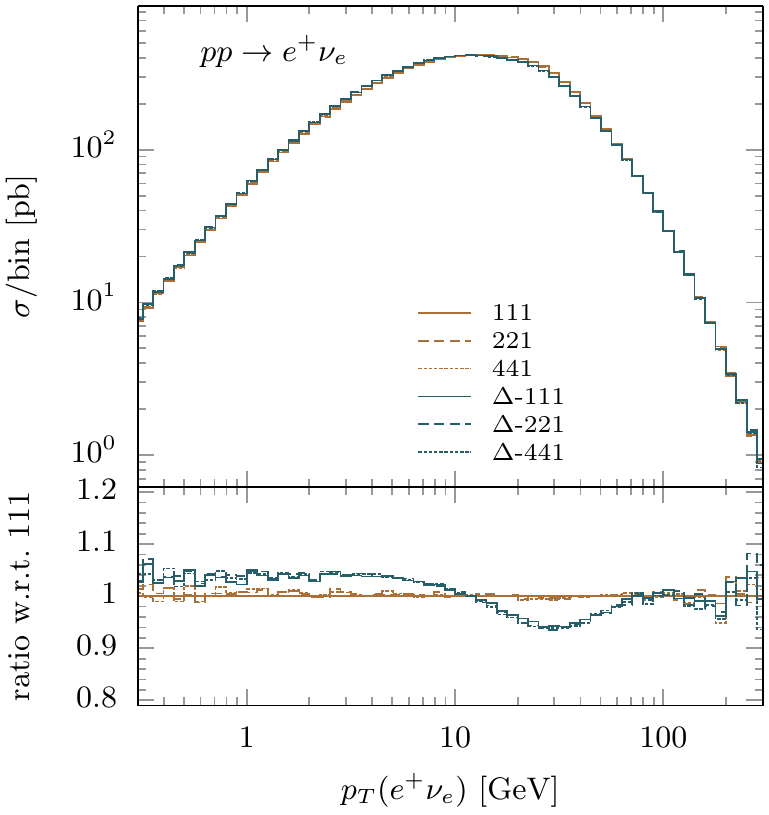}
\caption{\label{fig:DY} 
Transverse momentum of the $e^+e^-$ pair (left panel) and 
of the $e^+\nu_e$ pair (right panel), for the process in 
eqs.~(\ref{procee}) and~(\ref{procenu}), respectively. Both
MC@NLO (brown histograms) and $\aNLOD$ (blue histograms) results
are shown, for three choices of folding parameters.
}
  \end{center}
\end{figure}
We now turn to considering differential distributions. We point out
that, for each process, we have studied the behaviour of dozens of 
observables, with the largest numbers in the cases of the processes
with richer final states. For the sake of this paper, in view of the
limitations of space it entails, we have restricted ourselves to 
discussing the case of an observable which is very often used as a
test case in the context of matching procedures, namely the transverse
momentum of the set of particles that are present in the final state
at the Born level (for the process in eq.~(\ref{procWj}) such a quantity
is ill-defined; we use the $W$-hardest jet system instead, which is its
infrared-safe analogue). This observable has several appealing features.
For example, it constitutes a practical choice for the pull introduced
in eq.~(\ref{pulldef}); thus, it is very directly connected to the
mechanism of the reduction of negative-weight events which is central
to this work. Furthermore, by simply considering a sufficiently large
domain, one can separately assess the behaviour of the matched
predictions in the soft/collinear, intermediate, and hard regions.
This is helpful also because the theoretical systematics due to matching
(at least, in MC@NLO-type prescriptions) is mostly concentrated in
the intermediate region; therefore, such an observable is essentially
a worst-case scenario as far as this systematics is concerned. Indeed,
for the purpose of a comparison between MC@NLO and $\aNLOD$ predictions,
we have verified that the conclusions that can be drawn by studying
the transverse momentum of the Born system have a rather general
validity, in that by far and large they encompass the cases of the 
other observables that we have considered.

\begin{figure}[thb]
  \begin{center}
  \includegraphics[width=0.47\textwidth]{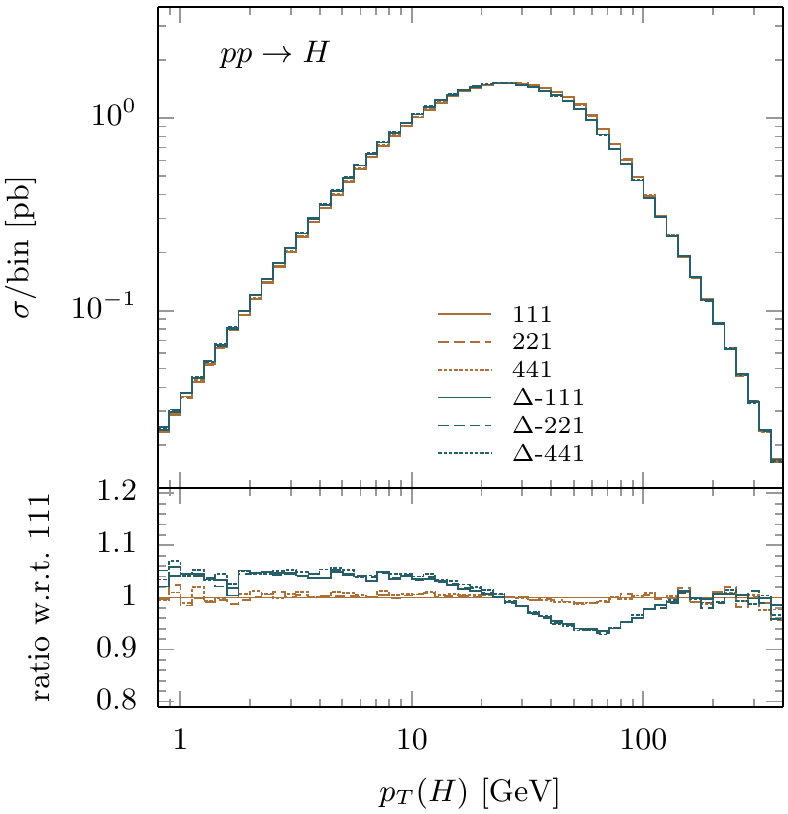}
$\phantom{a}$
  \includegraphics[width=0.47\textwidth]{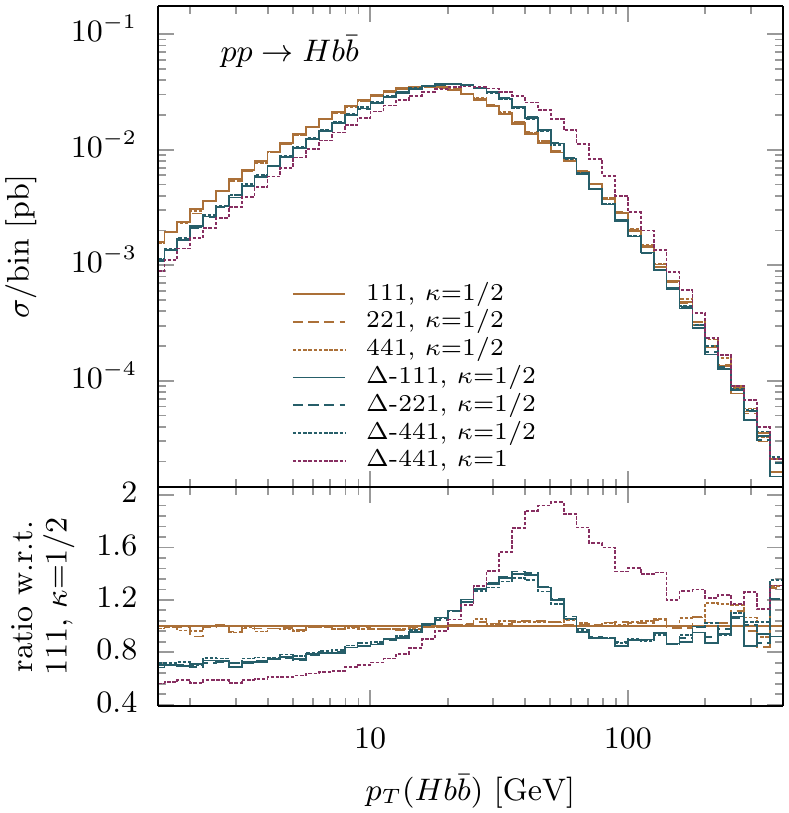}
\caption{\label{fig:Hgg}
As in fig.~\ref{fig:DY}, for the transverse momentum of the Higgs
(left panel) and of the $Hb\bb$ system (right panel), for the processes
in eqs.~(\ref{procH}) and~(\ref{procHbb}), respectively. The latter
displays an additional $\aNLOD$ curve (red dotted); see the text
for details, in particular eq.~(\ref{kffdef}) for the definition
of the parameter $\kappa$.
}
  \end{center}
\end{figure}

\begin{figure}[thb]
  \begin{center}
  \includegraphics[width=0.47\textwidth]{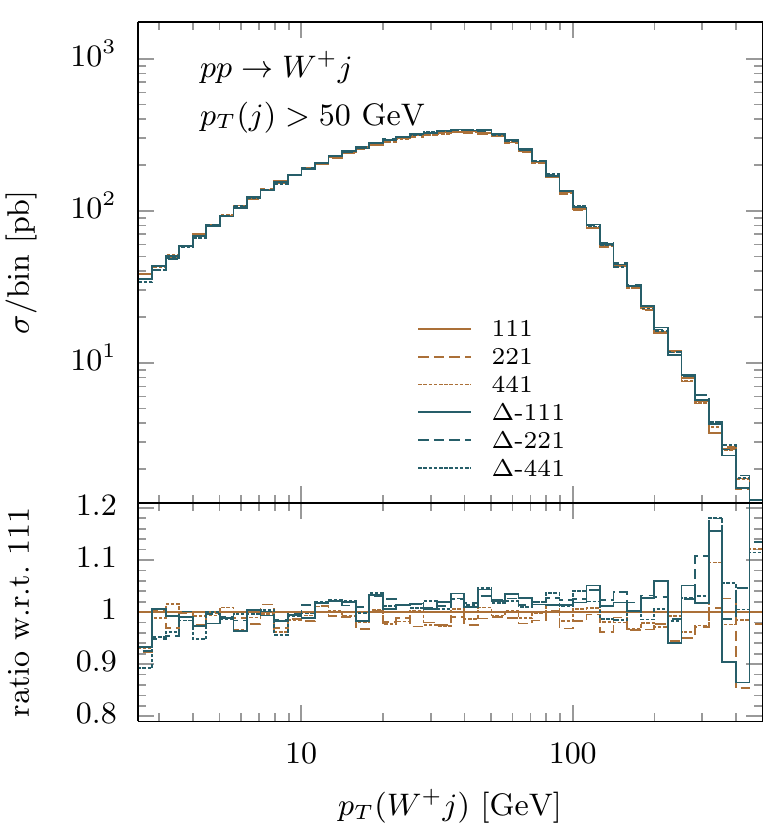}
$\phantom{a}$
  \includegraphics[width=0.47\textwidth]{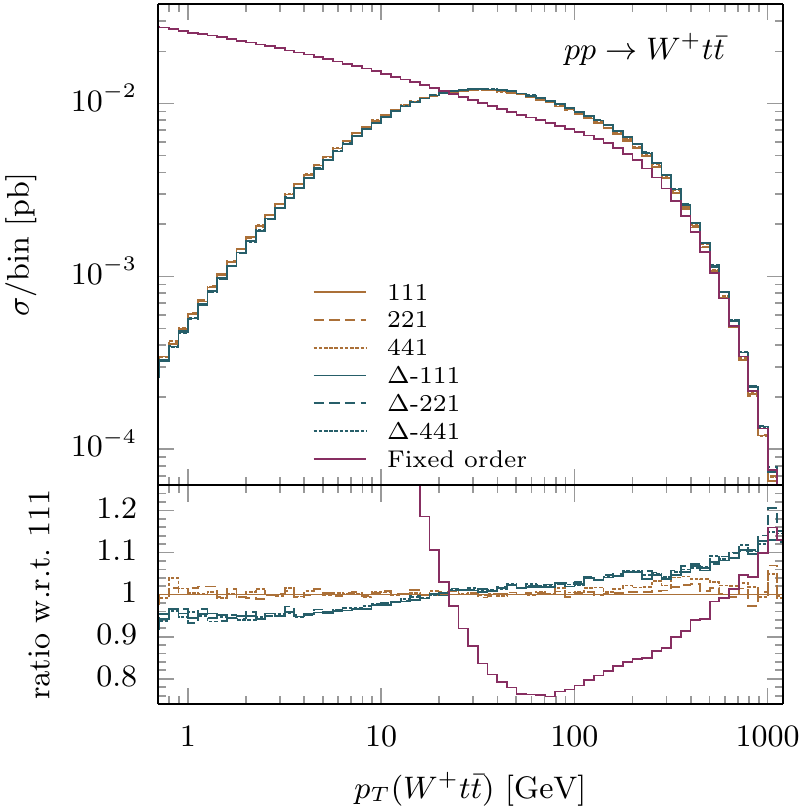}
\caption{\label{fig:W} 
As in fig.~\ref{fig:DY}, for the transverse momentum of the $W$-hardest jet
system (left panel) and of the $W^+t\bt$ system (right panel), for the processes
in eqs.~(\ref{procWj}) and~(\ref{procWtt}), respectively. In the right panel,
the fixed-order result (red solid) is also shown.
}
  \end{center}
\end{figure}

Our results are displayed in figs.~\ref{fig:DY}--\ref{fig:W} for the
processes in eqs.~(\ref{procee})--(\ref{procWtt}); the case of
$t\bt$ production will be dealt with later (see sect.~\ref{sec:ttb}).
These figures feature two panels, each relevant to a specific process,
composed of a main frame and a lower inset, and all have the same layout. 
In the main frame there are six histograms, three of which are relevant 
to MC@NLO predictions (brown lines) and the other three to $\aNLOD$ 
predictions (blue lines). The three results in each set differ from 
each other by the choice of the folding parameters: 
\mbox{$(n_\xi,n_y,n_\varphi)=(1,1,1)$} (i.e.~no folding; solid),
\mbox{$(n_\xi,n_y,n_\varphi)=(2,2,1)$} (dashed), and
\mbox{$(n_\xi,n_y,n_\varphi)=(4,4,1)$} (dotted).
Thus, the six curves in the main frames correspond to the six results
associated with each process in table~\ref{tab:procs}. The lower insets 
of the figures show the bin-by-bin ratios of each of the histograms that 
appear in the main frames, over the histogram relevant to the corresponding
MC@NLO result with no folding; the same plotting patterns as in the main 
frames are employed.

In fig.~\ref{fig:DY} we present the predictions for the $Z$- (left panel)
and $W^+$-mediated (right panel) first-family lepton pair production
of eqs.~(\ref{procee}) and~(\ref{procenu}); not surprisingly, they
look very similar to each other, and the only reason to show both of 
them is their prominence for both SM measurements and new-physics 
searches at the LHC. The MC@NLO and $\aNLOD$ results are within
$\pm 5$\% of each other in the whole range considered. They coincide
in absolute value at large $\pt$, in keeping with one of the defining
features of the MC@NLO matching, namely that in hard regions both the
shape and the normalisation must coincide, up to small shower effects, with 
the underlying fixed-order NLO prediction. This shows that such a feature
is also exhibited by $\aNLOD$, as is expected by construction. At small
transverse momenta the MC@NLO and $\aNLOD$ results have the same shapes,
which in the case of the former matching prescription is known to coincide
with that of the underlying parton-shower MC; again, this shows that this
property holds for $\aNLOD$ as well. As was anticipated, the differences
between MC@NLO and $\aNLOD$ are visible in the intermediate-$\pt$ region,
and must be attributed to matching systematics; in fact, they are of the
same order as those one would obtain e.g.~by varying the $f_\alpha$
parameters that control the starting scale(s) of the parton showers;
we shall give in sect.~\ref{sec:ttb} an explicit example on this point.
We finally remark that, for both of the matching prescriptions, the
differences between the results obtained with different folding parameters
are statistically compatible with zero, which confirms the nature of the
folding as a technical tool which has no impact on physics predictions.

In fig.~\ref{fig:Hgg} we show the results for production processes
that feature an SM Higgs -- single-inclusive (eq.~(\ref{procH})) in the 
left panel, and in association with a $b\bb$ pair (eq.~(\ref{procHbb}))
in the right panel. As far as single-inclusive production is concerned,
the same comments as for the lepton-pair processes of fig.~\ref{fig:DY} 
can be repeated verbatim. While this fact could have been expected, it
is nevertheless an excellent test of the $\aNLOD$ machinery. In fact, 
the production of lepton pairs is dominated by $q\bq$ annihilation, that
of Higgs by $gg$ fusion; as a consequence of this, the structures
of the $\Delta$ factors relevant to the two cases are very different
from each other.

The situation visibly changes for $Hb\bb$ production, which is 
a process notoriously affected by very large systematics (see 
e.g.~ref.~\cite{deFlorian:2016spz}). 
Thus, although the qualitative pattern of the MC@NLO vs $\aNLOD$
comparison is similar to that of the other processes considered so far,
in absolute value the differences are larger, up to $\pm 30$\%. 
We observe that the $\aNLOD$ results are harder than the MC@NLO ones
(i.e.~the peak of the distribution is at larger $\pt$ values). A by-product
of this fact is that the transition between the peak and the tail
regimes appears to be more regular for $\aNLOD$ than for MC@NLO -- 
in the latter case, the histograms are flatter in the region
\mbox{$20~\gev\lesssim\pt\lesssim 70~\gev$}. We also point out that, 
following ref.~\cite{Wiesemann:2014ioa}, $Hb\bb$ production has 
been simulated with the parameters of eq.~(\ref{kffHbb}). Interestingly,
in the case of MC@NLO, the choice of eq.~(\ref{kffval}) in conjunction 
with the reference scale of eq.~(\ref{Rdef}) (i.e.~a combination that had
not been considered in ref.~\cite{Wiesemann:2014ioa}) leads, for some 
observables, to a pathological behaviour -- the distribution and the
large fraction of negative-weight events render their cancellation
a very difficult task, so that the results are always affected by 
dominant statistical uncertainties. It is therefore reassuring that
this is {\em not} the case for $\aNLOD$: in fig.~\ref{fig:Hgg} we
show the result for this matching prescription (and one choice of
folding parameters, the others being statistically identical to it)
obtained with eq.~(\ref{kffval}) (red dotted histogram); this result
is as well behaved as those obtained with eq.~(\ref{kffHbb}). 
It is clear that large matching uncertainties remain, but this appears
to be characteristic of this production process at this perturbative
order; if anything, the present $\aNLOD$ systematics are smaller than
their MC@NLO counterparts of ref.~\cite{Wiesemann:2014ioa} (see in
particular fig.~5 there).

We finally turn to discussing $W^+$ production in association with
a light jet (eq.~(\ref{procWj}), left panel of fig.~\ref{fig:W}) and 
with a $t\bt$ pair (eq.~(\ref{procWtt}), right panel of fig.~\ref{fig:W}).
In the case of the former process, a \mbox{$\pT\ge 50~\gev$} cut on the 
$R=0.5$, anti-$\kt$ hardest jet is imposed. The MC@NLO and $\aNLOD$
predictions for $W^+j$ production are rather close to each other;
differences are generally smaller than $\pm 5$\%. We note that 
the ratio plots are flatter than those relevant to the processes
considered so far; this is in part because for the present observable
the separation between the various regimes is not as clear-cut as in 
the other cases (owing to the jet-$\pt$ cut, one is more inclusive
here)\footnote{We remark that $W^+j$ production, as well as any other
process that features jets at the Born level, does not require any
special treatment as far as scale assignments in $\aNLOD$ are concerned.
The procedure presented in sect.~\ref{sec:LHEscales} guarantees that no
strong hierarchy is created. Note that, in order for an NLO-accurate
generation to be sensible, Born-level configurations must feature scales
of comparable hardness.}. As far as $W^+t\bt$ production is concerned, the 
differences between the MC@NLO and $\aNLOD$ results are again quite small. 
However, at variance with what happens in the other processes, the $\aNLOD$ 
large-$\pt$ tail is harder than the MC@NLO one. In order to investigate
this point further, in fig.~\ref{fig:W} we also display the fixed-order (FO)
result (red solid histogram). This shows that the matched predictions
are significantly different w.r.t.~the FO one for up to very large
transverse momenta; in other words, the asymptotic regime (where 
MC@NLO-type and FO result are expected to coincide with each other) 
is approached in a very slow manner; essentially, one is not yet there
at the rightmost end of the range shown in fig.~\ref{fig:W}. Furthermore, 
the extreme steepness of the distribution at large $\pt$ implies that even 
small parton-shower effects might induce visible bin migrations; this is a 
timely reminder of the fact that, in this region, the predictions we are 
considering are LO-accurate in perturbation theory. We have verified
that, by plotting this distribution at the level of the hard events,
the MC@NLO and $\aNLOD$ results are on top of each other, and on top
of the FO one; thus, the differences between the former two predictions 
are indeed stemming from different shower-scale assignments.

\subsection{A closer look at $t\bt$ production\label{sec:ttb}}
\begin{figure}[thb]
  \begin{center}
  \includegraphics[width=0.55\textwidth]{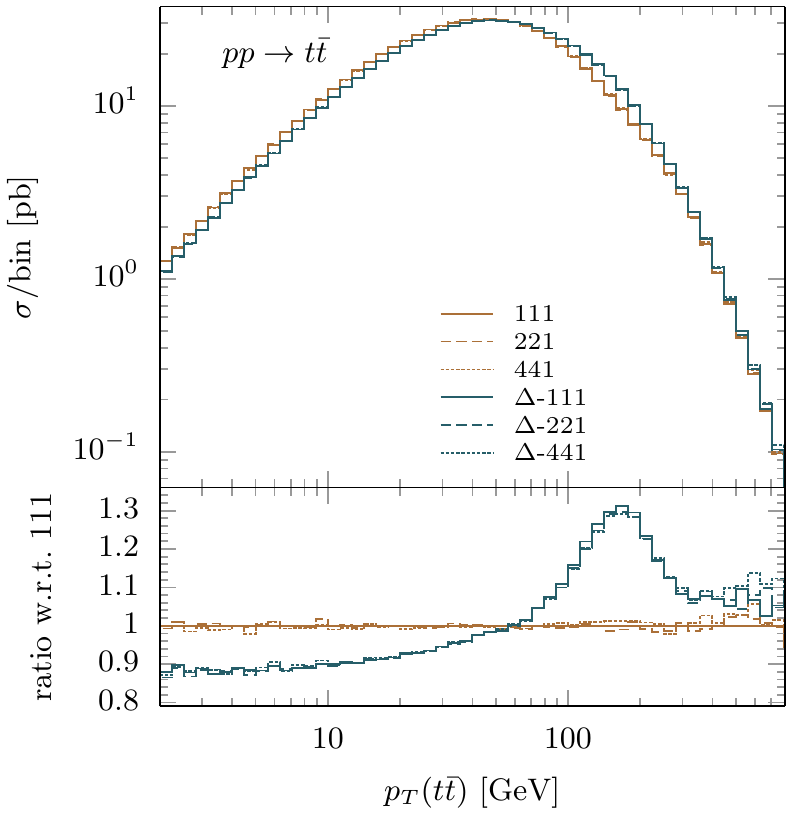}
\caption{\label{fig:ttpt} 
As in fig.~\ref{fig:DY}, for the transverse momentum of the $t\bt$
pair, for the process in eq.~(\ref{proctt}).
}
  \end{center}
\end{figure}

\begin{figure}[thb]
  \begin{center}
  \includegraphics[width=0.8\textwidth]{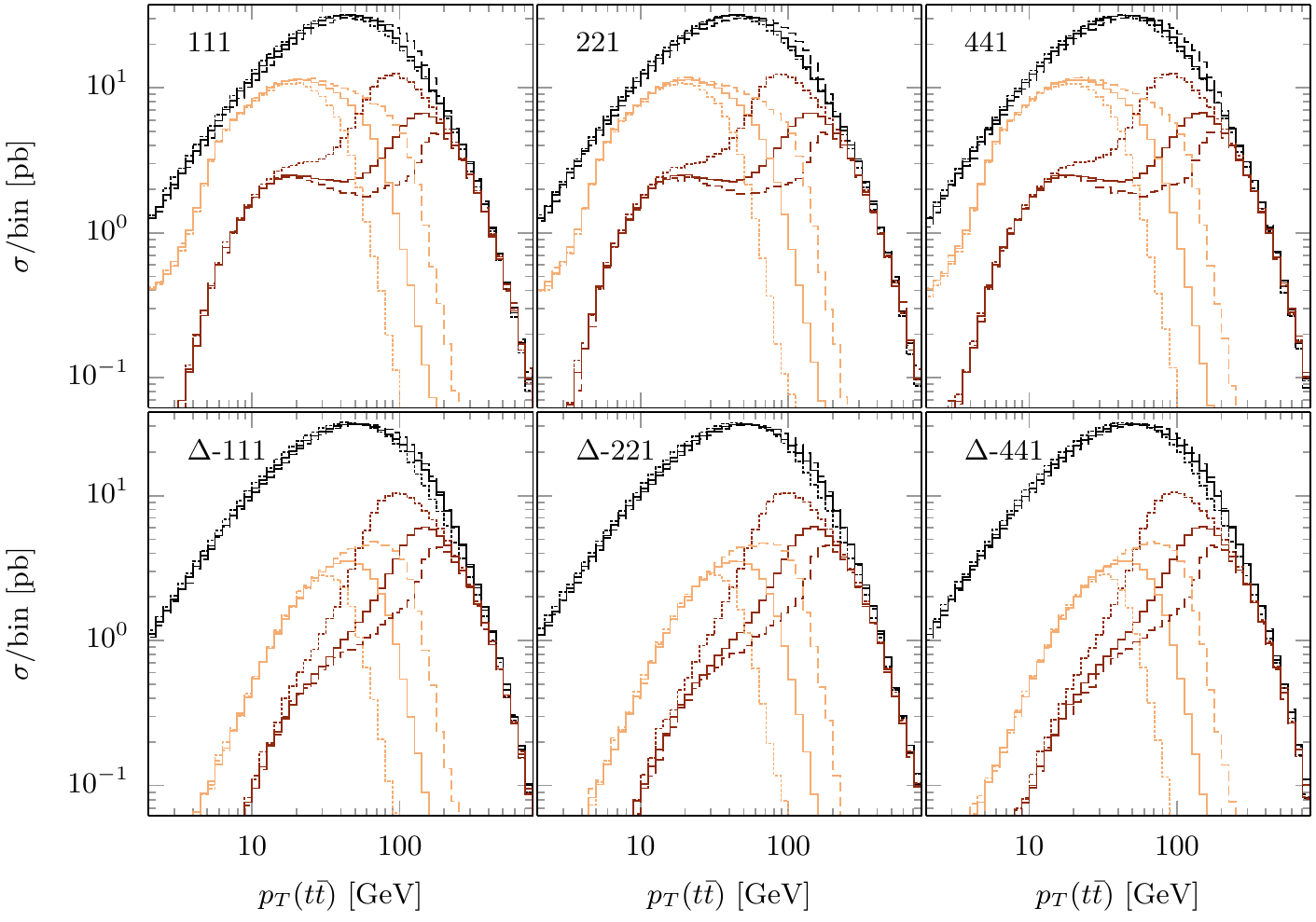}
\caption{\label{fig:ttptLHE} 
Transverse momentum of the $t\bt$ pair, after parton showers (black
histograms), and before parton showers for positive-weight (dark brown 
histograms) and negative-weight (light brown histograms) events;
see the text for details.
}
  \end{center}
\end{figure}

In this section we discuss in more details the case of $t\bt$
hadroproduction (eq.~(\ref{proctt})); this will give us the
opportunity to re-consider the arguments of sect.~\ref{sec:negw}
in light of a specific example. We start by considering the analogue 
of the observables displayed in figs.~\ref{fig:DY}--\ref{fig:W}, namely
the transverse momentum of the $t\bt$ pair. This is shown in
fig.~\ref{fig:ttpt}, whose layout is identical to that of the
figures considered so far. The conclusions are also qualitatively 
fairly similar to those relevant to the other processes. We note that 
the differences between the MC@NLO and $\aNLOD$ predictions are larger
here in the intermediate region; we anticipate (see fig.~\ref{fig:ttptLHE})
that this is not the signal of a genuine discrepancy between the two 
prescriptions, but it rather reflects a matching uncertainty that affects 
both matching techniques in a similar manner. The large-$\pt$ region also
exhibits some of the features relevant to $W^+t\bt$ production that we
have described in fig.~\ref{fig:W}; they are however milder here, owing
to the distribution being less steep and the final-state less massive
than in the previous case (thus, the asymptotic regime occurs at $\pt$
values relatively smaller than in the case of $W^+t\bt$ production).

We now study how the predictions after parton showers
(e.g.~those in fig.~\ref{fig:ttpt}) are related to the hard events
they stem from; the reader must bear in mind that, as always in the
context of MC@NLO-type matchings, the latter quantities are meaningful
only in a technical, as opposed to a physical, sense. In order to do
so, on top of considering the results obtained with the parameters
in eq.~(\ref{kffval}), we also use what follows:
\beqn
&&\kappa=1\,,\;\;\;\;\;\;
\hat{f}_1=0.1\,,\;\;\;\;\;\;\;
\hat{f}_2=0.55\,,
\label{kffvalL}
\\
&&\kappa=1\,,\;\;\;\;\;\;
\hat{f}_1=0.55\,,\;\;\;\;\;\;
\hat{f}_2=1\,.
\label{kffvalU}
\eeqn
We call the settings of eqs.~(\ref{kffval}), (\ref{kffvalL}),
and~(\ref{kffvalU}) the {\em baseline}, {\em lower}, and {\em upper} 
choices respectively, and abbreviate these names where necessary as 
{\em b}, {\em l}, and {\em u}. We stress that this notation has no
phenomenological implications. In other words, we do not claim that
the baseline choice should be regarded as the default option when
a comparison to data is performed; it is simply the choice of settings
which is formally closest to the one usually adopted in \aNLOs\ MC@NLO
simulations. Whether it is the baseline, the lower, or the upper settings
(or something else altogether) that are best for phenomenological $\aNLOD$ 
applications is an open question, which is beyond the scope of the present 
paper.

In fig.~\ref{fig:ttptLHE} we consider again the transverse momentum
of the $t\bt$ pair. There are six panels in the figure, and all of them
have the same layout. The black histograms are the MC@NLO (upper row)
and $\aNLOD$ (lower row) results; by definition, they are obtained
after parton showers. The brown histograms represent results at
the hard-event level, i.e.~before parton showers; the dark-brown (peaking
at relatively larger $\pt$'s) and light-brown (peaking at relatively smaller
$\pt$'s) ones are associated with positive-weight and negative-weight
events, respectively; in order to display them together, the latter are 
plotted by flipping their signs. Note that only $\clH$ events contribute 
to the hard-event histograms, since for $\clS$ events one has 
$\pt(t\bt)=0$, which is outside the domain of fig.~\ref{fig:ttptLHE}.
For each type of histogram (after parton showers and before parton showers
with event weights of either sign) there are three curves; the dotted,
solid, and dashed ones are obtained with the lower (eq.~(\ref{kffvalL})), 
baseline (eq.~(\ref{kffval})), and upper (eq.~(\ref{kffvalU})) settings,
respectively. It follows that the solid black histograms are the same
predictions as those already shown in fig.~\ref{fig:ttpt} as blue curves. 
The three panels in each of the two rows of fig.~\ref{fig:ttptLHE} differ 
by the choice of the folding parameters.

\begin{figure}[thb]
  \begin{center}
  \includegraphics[width=0.7\textwidth]{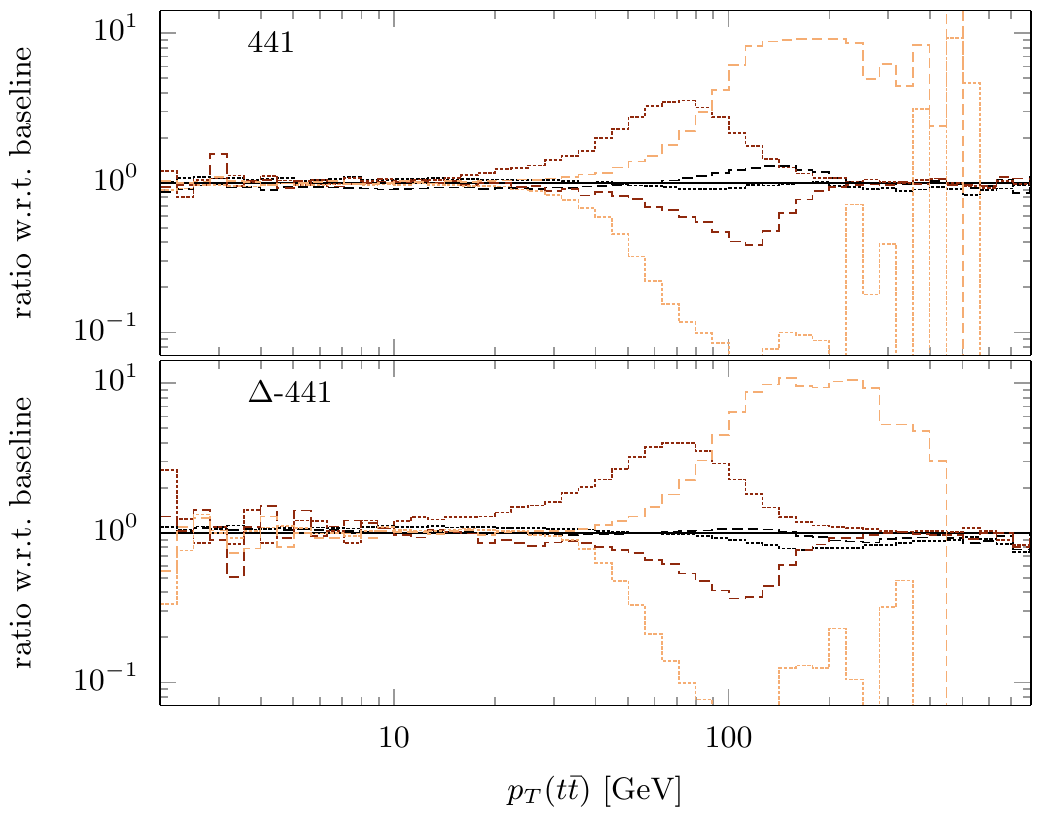}
\caption{\label{fig:ttptLHErat} 
Ratios of the upper and lower predictions over the baseline ones, for
the two rightmost panels of the two rows in fig.~\ref{fig:ttptLHE}.
}
  \end{center}
\end{figure}
There are several pieces of information that can be extracted from 
fig.~\ref{fig:ttptLHE}. We start by observing that the three panels
on each row are essentially indistinguishable from one another. This
is what we expect, since they differ by the choice of the folding
parameters; such a choice must induce differences statistically compatible
with zero for physical distributions (black histograms), and must not
affect $\clH$-event distributions (brown histograms) -- note that the
latter statement would not be true in the case of $\clS$ events, as 
we shall document below. The choice of the $\hat{f}_\alpha$ parameters
gives a measure of the matching systematics after parton showers;
we note that such systematics (identified as the cumulative difference 
among the three black histograms that appear on any given panel) are very 
similar, but not identical, in MC@NLO and $\aNLOD$; this confirms that,
for precision studies, $\aNLOD$ matching uncertainties are best assessed
independently of the corresponding MC@NLO ones. As was anticipated, 
the MC@NLO and $\aNLOD$ systematics emerging from fig.~\ref{fig:ttptLHE} 
are consistent with the differences between the MC@NLO and $\aNLOD$
predictions of fig.~\ref{fig:ttpt} (i.e.~at most a relative ~$30$\% effect, 
but typically smaller than that). At variance with the physical results, 
those at the hard-event level (brown histograms) exhibit a very strong 
dependence on the $\hat{f}_\alpha$ parameters (up to a relative factor 
of $5$ ($10$) w.r.t.~the central results for the dark brown (light brown) 
curves): this need not be surprising, owing to their unphysical nature.
However, such a large dependence offers one the possibility of a better 
understanding of the interplay between positive weights, negative
weights, and scale assignments (whose general features have been 
introduced and discussed in sect.~\ref{sec:negw}). In order to give a
more quantitative idea of the sizes of the effects we have just discussed,
we present in fig.~\ref{fig:ttptLHErat} the ratios of the upper and lower 
predictions over the corresponding baseline ones, for the rightmost panels
in the two rows of fig.~\ref{fig:ttptLHE}, that correspond to the 
$(4,4,1)$ choice of the folding parameters -- the pattern emerging from
other folding-parameter choices is the same as that of 
fig.~\ref{fig:ttptLHErat}.

In order to proceed, we note preliminarily that the hard-event distributions
in fig.~\ref{fig:ttptLHE} are nothing but the distribution in $\pt(t\bt)$
as is predicted solely by the short-distance cross section $d\sigmaH$
(for MC@NLO, see eq.~(\ref{genMCatNLO})) and $d\sigmaHD$ (for $\aNLOD$, 
see eq.~(\ref{genMCatNLOD})); one simply plots the positive and negative 
contributions separately. This follows from the fact that by not performing 
parton showers one leaves the parton-level kinematical configurations 
unchanged. Furthermore, we remark that the value of $\pt(t\bt)$ 
is a good estimate of the stopping scale relevant
to the Sudakov form factor that dominates in the construction of $\Delta$
(i.e.~the smallest one); as far as the corresponding starting scale is 
concerned, its value will be in the range 
\mbox{$(\kappa\hat{f}_1\mu_t,\kappa\hat{f}_2\mu_t)$} (see eqs.~(\ref{f1f2mat})
and~(\ref{startSdef})), with $\mu_t$ predominantly of the order of the top 
mass; on average, the starting scale will be roughly equal to
\mbox{$\kappa(\hat{f}_1+\hat{f}_2)\mu_t/2$}.

As the comparison of eq.~(\ref{xsecH}) with eq.~(\ref{xsecHD}) shows,
we have:
\beq
d\sigmaHD=d\sigmaH\Delta\equiv
\big(d\sigmaNLOE-d\sigmaMC\big)\Delta\,.
\eeq
One of the implications of what has been said about the stopping and 
starting scales is that \mbox{$\Delta\to 0$} for small $\pt(t\bt)$.
Thus, the cross section in this region must be much smaller
in $\aNLOD$ than in MC@NLO -- this is what one sees by comparing
the brown histograms in the lower row of fig.~\ref{fig:ttptLHE} with
the corresponding upper-row ones, the differences being increasingly
large as one moves towards $\pt(t\bt)=0$. This mechanism is the 
anticipated suppression of \nclo\ events in $\aNLOD$ w.r.t.~those
in MC@NLO; as is expected, positive-weight events are suppressed 
as well, but this does not lead to a degradation of the statistical
accuracy as is explained just before sect.~\ref{sec:Dcostr}.

\begin{figure}[thb]
  \begin{center}
  \includegraphics[width=0.8\textwidth]{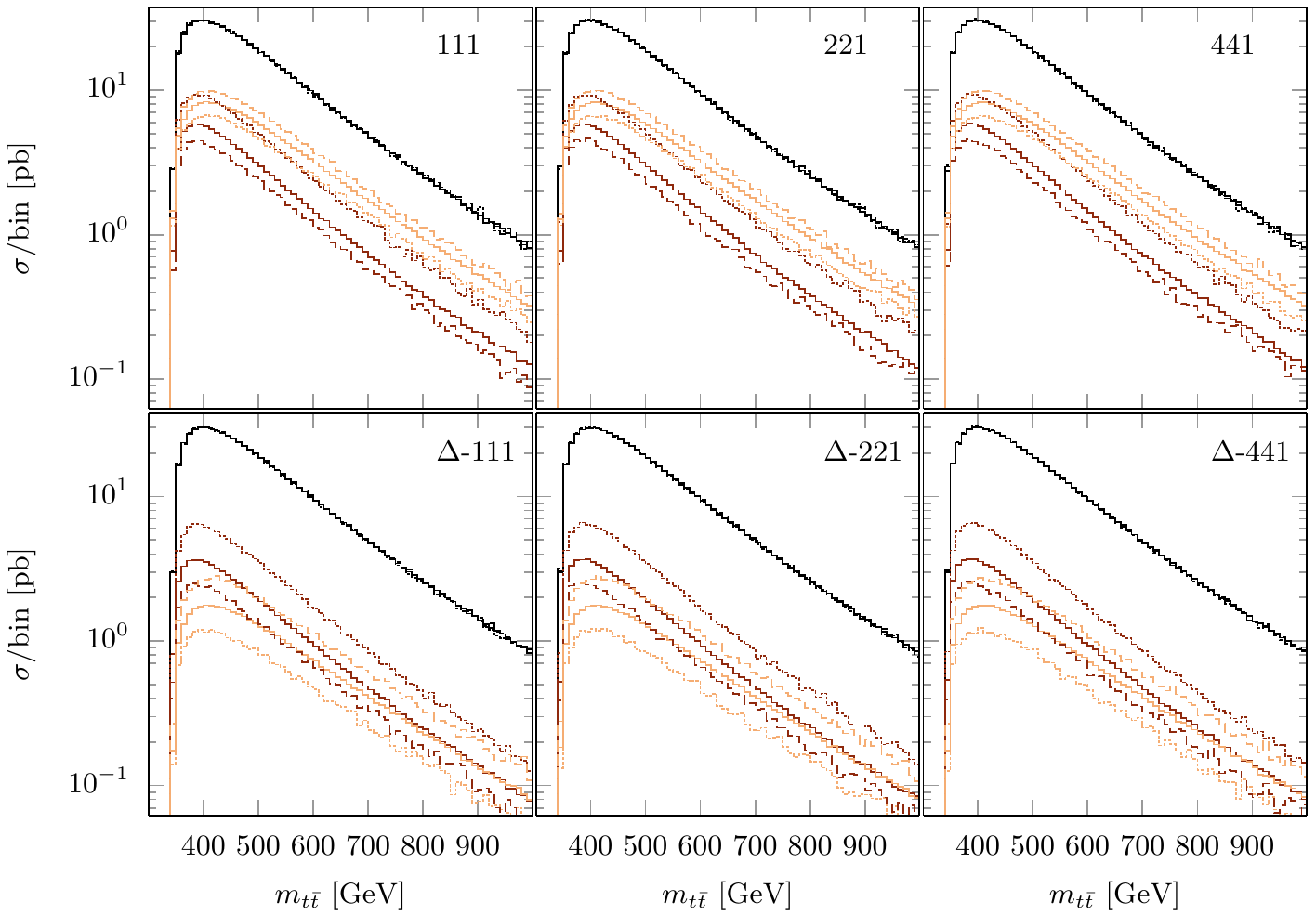}
\caption{\label{fig:ttMLHEH} 
As in fig.~\ref{fig:ttptLHE} for the invariant mass of the $t\bt$ pair.
For hard events before parton shower, only $\clH$ ones are shown.
}
  \end{center}
\end{figure}
The comparisons between the dotted, solid, and dashed brown histograms
relevant to the same type of events (i.e.~with positive or negative
weights) show that the dependence on the choice of the lower, baseline,
and upper settings is negligible at small $\pt(t\bt)$, but becomes
visible at larger transverse momenta; this is a feature present in
both MC@NLO and $\aNLOD$ predictions, and can be understood as follows.
The values of the $\hat{f}_\alpha$ parameters affect the starting
scales, but not the stopping scale; accordingly, there is a corresponding 
dependence of the $\Delta$ factor. However, such a dependence is never
particularly large, being roughly equal to the ratio of the Sudakovs
that dominate in the $\Delta$ factor, computed at the different starting
scales (see eq.~(\ref{sudscales})); since the latter are typically large, 
these Sudakovs are then of order one, and so is their ratio. Furthermore, 
as the stopping scale tends towards zero, the values of $\Delta$ become 
quite small, so that $d\sigmaHD$ is vanishingly small as well; this implies
that the modest dependence of $\Delta$ on $\hat{f}_\alpha$ hardly changes
the cross section (in other words, a large suppression factor is as 
effective as the same suppression factor multiplied by a number of order one).
Things do change at larger $\pt(t\bt)$: here, the situation is complicated 
by the fact that the dependence of the short-distance cross section on 
$\hat{f}_\alpha$ is due not only to the $\Delta$ factor, but also to the
function $D$ that appears in the MC counterterms (see eq.~(\ref{MCcnts});
note that $D=1$ for small transverse momenta). For any given $\pt(t\bt)$,
the $\hat{f}_\alpha$ dependencies of these quantities result in
competing effects in the case of $\aNLOD$, while only the latter is
present in MC@NLO (owing to the absence of $\Delta$ there): larger 
$\hat{f}_\alpha$ give smaller $\Delta$ (i.e.~more suppression 
of both positive- and negative-weight events in $\aNLOD$) and larger 
$D$ values (i.e.~larger MC counterterms and therefore more negative-weight 
events in both $\aNLOD$ and MC@NLO; see eq.~(\ref{Ddef})). Which of the two 
mechanisms is dominant in $\aNLOD$ depends on the kinematical regions one 
is interested in. In general, given the functional form of the Sudakov form 
factors, and in particular the way in which they approach the value of one,
as $\pt(t\bt)$ increases the effect of the $\Delta$ factor is 
subdominant w.r.t.~that of the $D$ function, {\em a fortiori} in the case
of MC@NLO. This appears to imply that, perhaps contrary to naive 
expectations, smaller $\hat{f}_\alpha$ are more effective in reducing 
negative-weight events not only in MC@NLO, but in $\aNLOD$ as well.
However, it is important to stress that, while the mechanisms described 
here have a general validity, their relative balance in $\aNLOD$ is delicate, 
and dependent on the process. As far as $t\bt$ production is concerned, 
we shall later document how the conclusion above can be quantified.
\begin{figure}[thb]
  \begin{center}
  \includegraphics[width=0.8\textwidth]{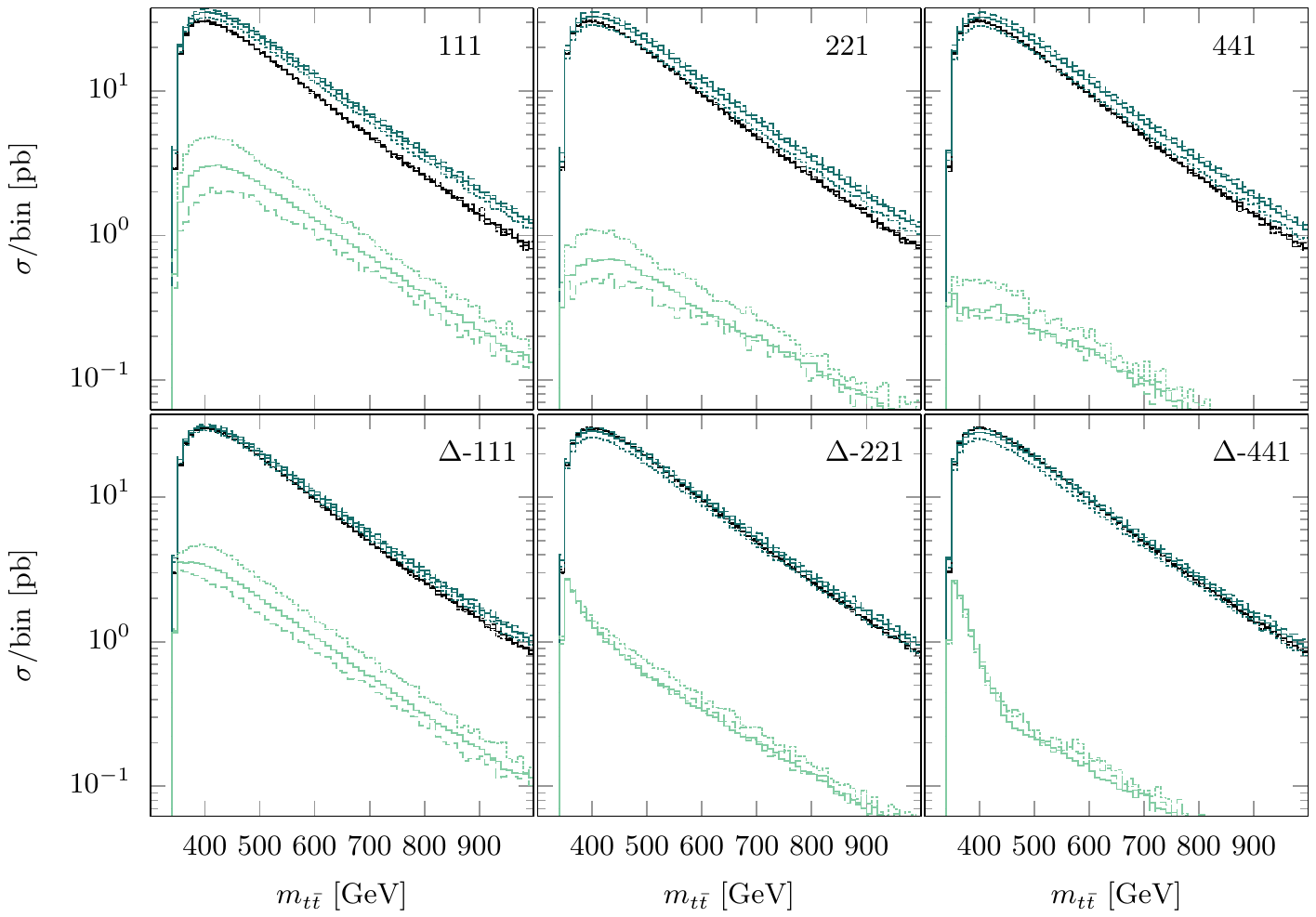}
\caption{\label{fig:ttMLHES} 
As in fig.~\ref{fig:ttMLHEH}, for $\clS$ events rather than for $\clH$ events.
}
  \end{center}
\end{figure}

In terms of the classification of negative-weight events given in
sect.~\ref{sec:negw}, we can recapitulate the discussion above by 
saying that the factor $\Delta$ primarily affects the number
of \nclo\ events and, to a lesser extent, that of \nclt\ events.
As far as the $D$ function is concerned, it mostly affects the
number of \nclt\ events, and only marginally that of \nclo\
events. Thus, in $\aNLOD$ by decreasing (increasing) the $\hat{f}_\alpha$ 
values one eliminates less (more) \nclo\ events, while simultaneously
eliminating more (less) \nclt\ events. In MC@NLO, only the \nclt\
events are affected.

\begin{figure}
  \begin{center}
  \includegraphics[width=0.99\textwidth]{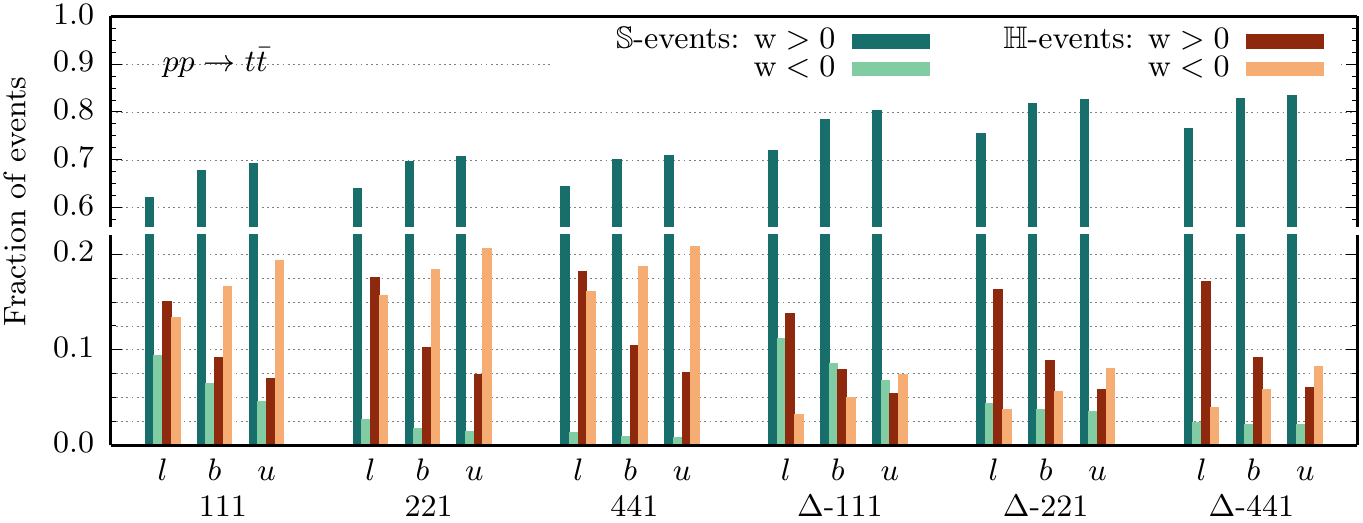}
\caption{\label{fig:ttNev} 
Fractions of positive- and negative-weight events, for both $\clS$ and
$\clH$ events, in $t\bt$ production.
}
  \end{center}
\end{figure}
Before summarising our findings, we present plots which are the analogues
of those in fig.~\ref{fig:ttptLHE}, but are relevant to the invariant
mass of the $t\bt$ pair rather than to $\pt(t\bt)$. The results
are shown in fig.~\ref{fig:ttMLHEH} and fig.~\ref{fig:ttMLHES}.
The layout of fig.~\ref{fig:ttMLHEH} is exactly the same as that
of fig.~\ref{fig:ttptLHE}, namely predictions after parton shower
(black histograms) and before parton shower for $\clH$ events
(brown histograms).
In fig.~\ref{fig:ttMLHES} we display the same physical results
as in fig.~\ref{fig:ttMLHEH}, but the distributions at the hard-event
level are obtained with $\clS$ events, rather than with $\clH$ events --
at variance with the case of the $t\bt$ pair transverse momentum, $\clS$-event
contributions are non-trivial in the whole $m_{t\bt}$ range, and are
in fact the dominant ones. Positive- and negative-weight $\clS$-event
distributions are displayed as dark- (relatively larger cross sections)
and light-green (relatively smaller cross sections) histograms, respectively.
The interpretation of figs.~\ref{fig:ttMLHEH} and~\ref{fig:ttMLHES}
is not as easy as that of fig.~\ref{fig:ttptLHE}, owing to the fact
that the pair invariant mass is not a pull variable. Therefore, all
of the effects that we have discussed for $\pt(t\bt)$ simultaneously
contribute to the same $m_{t\bt}$ value, and are impossible to
disentangle, although for example the pattern associated with the
choices of the $\hat{f}_\alpha$ parameters can still be seen.
Furthermore, fig.~\ref{fig:ttMLHES} presents one with the direct
evidence of the effects of the folding, which cannot be seen in the 
$\clH$-event distributions of figs.~\ref{fig:ttptLHE} and~\ref{fig:ttMLHEH},
in particular for what concerns the reduction of \nclth\ events.

We conclude this section by presenting a summary of the situation of
the fractions of positive- and negative-weight events in $t\bt$ production,
fully inclusive over all kinematical variables. In other words, we simply
count the number of positive and negative weights present in the various
LHE files, and divide them by the total number of events; by using
the notation introduced in sect.~\ref{sec:intro}, these ratios are
therefore equal to $1-f$ and $f$, respectively, when $\clS$ and $\clH$
events are considered together. By keeping the $\clS$- and $\clH$-event
contributions separate from each other, we obtain the results shown
in fig.~\ref{fig:ttNev}. In that figure, there are three groups
of four bins for each of the three MC@NLO and $\aNLOD$ scenarios
(no folding, $221$ folding, and $441$ folding) that we have considered.
Each one of such groups corresponds to a choice of the $\hat{f}_\alpha$ 
parameters, and is labelled accordingly: lower choice ({\em l}, left), 
baseline choice ({\em b}, center), and upper choice ({\em u}, right). 
Finally, the four bins in each group give the fractions of the various 
events w.r.t.~the total number of events (i.e.~the sum of the contents of
the four bins is always equal to one); from left to right, positive-weight
$\clS$ events (dark green), negative-weight $\clS$ events (light green), 
positive-weight $\clH$ events (dark brown), and negative-weight $\clH$ 
events (light brown). In summary, the information contained in each of the
bins of fig.~\ref{fig:ttNev} is the phase-space integrated version of
that given, at the differential level, by the brown and green histograms
in figs.~\ref{fig:ttptLHE}--\ref{fig:ttMLHES}.
\begin{figure}
  \begin{center}
  \includegraphics[width=0.55\textwidth]{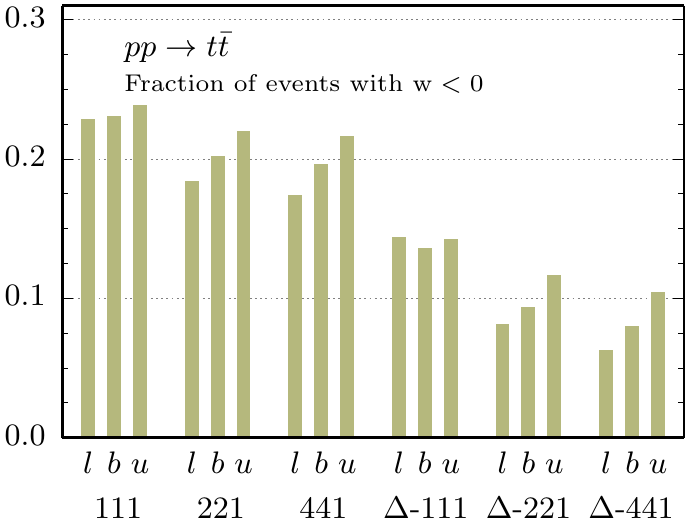}
\caption{\label{fig:ttNev2} 
Fractions of negative-weight events, in $t\bt$ production.
}
  \end{center}
\end{figure}
It should be noted that, by normalising all fractions
to the total number of events, in both MC@NLO and $\aNLOD$ matchings when
folding is applied both of the fractions relevant to $\clH$ events increase
(w.r.t.~the case of no folding), while either the negative- or the 
positive-weight fraction of $\clS$ events decreases; in non-pathological 
situations, it is the former that decreases and the latter that increases. 
We point out that this is just a by-product of the chosen normalisation, 
and that the short-distance cross sections have the behaviour already
described in sects.~\ref{sec:negw} and~\ref{sec:mcnloD}, namely that
the $\clH$-event cross section is unaffected by folding, while the negative
contributions to the $\clS$-event one are reduced. Furthermore, when $\clS$
and $\clH$ events are considered together, the total fraction of 
negative weights always decreases with folding (see fig.~\ref{fig:ttNev2}).
Figure~\ref{fig:ttNev} summarises what we have discussed so far.
For a given choice of the $\hat{f}_\alpha$ and folding parameters, we 
see a marked decrease in the fraction of negative-weight $\clH$ events, 
and a slight increase in that of negative-weight $\clS$ events, in
$\aNLOD$ w.r.t.~MC@NLO. For a given choice of the $\hat{f}_\alpha$ parameters,
in both MC@NLO and $\aNLOD$ by folding one visibly reduces the fraction
of negative-weight $\clS$ events. The overall fraction ($f$) of
negative-weight events that results from the various scenarios
is presented in fig.~\ref{fig:ttNev2}, which re-iterates the 
anticipated fact that folding is always advantageous. In conclusion, 
figs.~\ref{fig:ttNev} and~\ref{fig:ttNev2} confirm in more details
and for the specific case of $t\bt$ production the message that emerges
from table~\ref{tab:procs}: the reduction of the relative cost is 
achieved by moving from MC@NLO to $\aNLOD$, and by increasing the
amount of folding. The relative cost also depends on the parameters 
that control the assignments of the starting scales; although the 
latter might thus be used to further reduce it, this operation
must follow their tuning to data, and within the uncertainties
associated with such a tuning.

\section{Conclusions\label{sec:conc}}
In this paper we have introduced a new NLO-accurate matching 
prescription, that we have called $\aNLOD$, by starting from the 
MC@NLO cross section and by augmenting it with further elements 
of parton-shower MC origin, with the ultimate goal of reducing
the number of negative-weight events that are present in MC@NLO
simulations. We have achieved a practical implementation of the
$\aNLOD$ matching, by including it in the \aNLO\ framework
in conjunction with the \PYe\ MC. We have also taken this 
opportunity to implement in \aNLO\ the folding~\cite{Nason:2007vt}
technique, which induces a (further) reduction of negative-weight 
events in both $\aNLOD$ and MC@NLO. Finally, we have shown how the
$\aNLOD$ matching naturally results in the inclusion, in the hard
events eventually to be processed by the MC, of as many shower scales
as there are oriented colour connections; thus, there are one (in the 
case of quarks) or two (in the case of gluons) scales per external 
strongly-interacting particle. This is expected to give a better 
modelling w.r.t.~the one which is currently in use, based on a
single shower scale per event.

We have presented a systematic comparison of the MC@NLO and $\aNLOD$
predictions for seven processes in hadronic collisions at $\sqrt{S}=13$~TeV, 
which constitute a typical set of applications of NLO-matched computations
to LHC physics. The conclusion emerging from such a comparison is that
the combination of the $\aNLOD$ matching and of folding leads to a 
significant reduction of the number of negative weights that affect MC@NLO 
simulations.

Although the physical results we have presented are based on the
\PYe\ MC, as is the case for MC@NLO the $\aNLOD$ matching has a general
validity, and can be applied to any parton shower MC. However, one must
bear in mind that technically the structure of the implementation in
\aNLO\ of the $\aNLOD$ matching is more involved w.r.t.~that of MC@NLO,
since at variance with the latter it requires calling MC modules also 
during the event-generation phase. Therefore, in order to replicate
this structure in the case of another MC, the latter should be flexible
enough to allow for this (which should be the case for any modern
C$++$-based MC). Alternatively, the relevant modules should be extracted
from the MC and included in the \aNLO\ (or, in general, the matrix-element
provider) code. For completeness, we mention that such modules must
include the capability, for any given kinematical configuration, of
computing the Sudakov form factors that enter the definition of $\Delta$,
of returning the stopping shower scales, and of deciding whether the
given configuration is in an MC dead zone or not.

Reducing the number of negative-weight events is always advantageous
from a statistical viewpoint; an elementary quantitative analysis is given 
in sect.~\ref{sec:intro}. However, as far as the associated financial
costs are concerned, one must take into account that such a reduction 
is generally accompanied by an increase of the cost per event (see 
eq.~(\ref{addcost})); importantly, this is limited to the event-generation 
phase, and it stems from the longer running times necessary to achieve
the reduction. This is the case for both the folding (which affects both
MC@NLO and $\aNLOD$; the increase in running time is limited from above by 
the product of the folding parameters, and is usually about 75\% of that
number), and the $\Delta$ prescription itself (in our \PYe\ implementation, 
the running time increases by a factor 2--3 for all of the processes we
have considered, with the largest increases affecting the simplest ones). 
Neither of these aspects should be important in the balance costs/benefits 
in the context of full experimental simulations, while they deserve 
consideration for purely-theoretical studies. Both are liable to be 
improved by streamlining their underlying software structures.

We conclude by observing that while $\aNLOD$ originates from
the MC@NLO prescription, it is a novel matching technique, whose
implications must not be assumed to be identical to those of MC@NLO.
In particular, we have shown that the physics predictions for 
several hadroproduction processes that originate from MC@NLO and
$\aNLOD$ are fairly similar to each other. However, exactly because
they are not identical, the parameter settings that are optimal
for phenomenology studies are most likely not the same in the two
cases; this fact has to be taken into account when comparing
$\aNLOD$ results with experimental data. From the theoretical
viewpoint, on top of the differences inherent to the matching,
those stemming from the novel multi-scale structure of hard-event
files will also deserve careful consideration.

\section*{Acknowledgements}

We warmly thank Fabio Maltoni and Bryan Webber, for their comments
on this manuscript, Valentin Hirschi, for his help during the very 
early stages of this work, and Josh Bendavid, for his encouragement
and the many discussions we have had over the years.
SF is grateful to Bryan Webber, for a long collaboration on these
matters, and to the CERN TH division, for the hospitality during
the course of this work.
RF~and SP~are supported by the Swedish Research Council under 
contract number 2016-05996.
SP has received funding from the European Union's Horizon 2020
research and innovation program as part of the Marie Sk\l{}odowska-Curie
Innovative Training Network MCnetITN3 (grant agreement no. 722104).

\appendix
\section{The Sudakovs in \PYe\label{sec:PYesud}}
In this appendix we report on some of the technical details relevant
to the pre-tabulation of the \PYe\ Sudakov form factors, which we
have introduced in sect.~\ref{sec:impl}.

We start by observing that the Sudakovs are defined by means of
eq.~(\ref{suddef}); in view of this, it is convenient to construct the 
look-up tables for $\Delta_a^{(*)}$ rather than for $\Delta_a$, owing to 
the simpler functional dependence of the former w.r.t.~that of the latter.
The direct dependence of $\Delta_a^{(*)}$ on the colour line $\ell$
amounts to knowing whether the beginning and the end of $\ell$ are
in the initial or in the final state\footnote{The reader must bear in
mind that $a$ belongs to $\ell$.}. There are therefore four possible
cases, namely initial-initial, initial-final, final-initial, and final-final,
that we shorten as \II,  \IF,  \FI,  and \FF,  and call {\em types}; the
first (second) letter identifies the beginning (end) of the colour line.
We also have:
\beq
\Mdips=\{m_a,\mdip\}\,,
\eeq
where $m_a$ is the mass of particle $a$ according to the internal 
MC settings, and $\mdip$ is the generalised invariant mass of the pair 
colour-connected by $\ell$, defined as the (square root of the) r.h.s.~of
eq.~(\ref{Rkpdef}).
In practice, the dependence of Sudakovs on $m_a$ is essentially negligible
w.r.t.~the other dependencies; therefore, in our simulations we have set
the value of $m_a$ to a fixed value, equal to the \PYe\ default for
particle $a$. We also exploit the fact that:
\beq
\Delta_a^{(*)}\left(\qt;\Qt,\ell,\Mdips\right)=
\frac{\Delta_a^{(*)}\left(\qt;\QMt,\ell,\Mdips\right)}
{\Delta_a^{(*)}\left(\Qt;\QMt,\ell,\Mdips\right)}\,.
\label{sudscales}
\eeq
In other words, thanks to eq.~(\ref{sudscales}) we can obtain the functional
dependence of the Sudakov on both $\qt$ and $\Qt$ in terms of a single
parameter, provided that the value of $\QMt$ is sufficiently large.
We thus arrive at what follows:
\begin{itemize}
\item For each Sudakov type (\II,  \IF,  \FI, \FF) and particle identity 
($u$, $d$, $\ldots t$, $g$ -- Sudakovs for antiquarks are the same as
for quarks) we set up a look-up table. There are therefore 28 tables 
in total.
\item Each entry of a given table is the value of the corresponding 
$\Delta_a^{(*)}$ factor computed at pre-selected values of $\sqrt{\qt}$ 
and $\mdip$, called {\em nodal points}. In other words, a look-up table is 
a set of numbers associated with a two-dimensional grid, defined by the nodal 
points in the two-dimensional space spanned by $\sqrt{\qt}$ and $\mdip$.
\end{itemize}
The nodal points for both the $\sqrt{\qt}$ and $\mdip$ variables are
chosen by means of the same functional form, namely:
\beq
\log_{10}\left(\frac{r(j)}{1~{\rm GeV}}\right)=
r_0\left(\left(\frac{j}{j_{\max}}\right)^b-a\right),
\;\;\;\;\;\;\;\;
1\le j\le j_{\max}\,,
\label{node1}
\eeq
with $r_0$ and $a$ such that:
\beq
r(1)=r_{\min}\,,
\;\;\;\;\;\;\;\;\;\;
r(j_{\max})=r_{\max}\,.
\label{node2}
\eeq
In eqs.~(\ref{node1}) and~(\ref{node2}) the parameters $j_{\max}$, $b$,
$r_{\min}$, and $r_{\max}$ are fixed; they must be chosen so as to optimise
the distributions of the nodal points. For the simulations presented in this
paper we have used the following settings:
\beq
b=1\,,
\;\;\;\;\;\;\;\;
r_{\min}=1~{\rm GeV}\,,
\;\;\;\;\;\;\;\;
r_{\max}=7~{\rm TeV}\,,
\label{nodepar1}
\eeq
for both the $\sqrt{\qt}$ and $\mdip$ variables, and
\beq
j_{\max}=100~~{\rm for}~~\sqrt{\qt}
\,,
\;\;\;\;\;\;\;\;
j_{\max}=50~~{\rm for}~~\mdip\,.
\label{nodepar2}
\eeq
Equation~(\ref{node1}) is such that there is a higher density of nodal 
points towards the lower end of the range, $r_{\min}$; this takes into 
account the fact that Sudakovs vary faster for small values of $\qt$ 
and $\mdip$. In order to increase (decrease) such a density one must 
choose larger (smaller) values of $b$ (in any case, it is best to 
have $b=\ord(1)$).

We have constructed a program ({\tt gridsudgen.f}), and included it in the 
\aNLOs\ package, that loops over the Sudakov types and particle identities; 
for each of these, it calls the \PYe\ module (which we briefly describe below) 
that computes the Sudakov for all possible pairs of nodal points. 
These nodal values are then written in a new program ({\tt sudakov.f})
that includes interpolating instructions as well (in turn, these exploit 
the {\tt dfint} routine of the CERN libraries, that performs a double
linear interpolation). The interpolation is protected in the case of
inputs which are outside of the grid of nodal points. In particular:
{\em a)} if $\mdip$ is smaller than its minimal value or larger than
its maximal value, an error is returned and the run is stopped;
{\em b)} if $\qt$ is larger than its maximal value $\QMt$, the Sudakov is 
set equal to one; {\em c)} if $\qt$ is smaller than its minimal value
$\qmt$, the Sudakov is set either equal to zero if $\qt<(0.5~\gev)^2$
(with $0.5$~GeV taken to be the typical low-energy MC cutoff),
or equal to the value resulting from a linear interpolation between
zero and the value assumed by the Sudakov at \mbox{$\qt=\qmt$}, if 
\mbox{$(0.5~\gev)^2\le\qt<\qmt$}. Lest the interpolating {\tt dfint} 
routine actually extrapolates, we also set (see eq.~(\ref{node2}))
$\qmt=r_{\min}^2$ and $\QMt=r_{\max}^2$. Finally, {\tt sudakov.f} is 
automatically linked to \aNLOs\ when creating the executable relevant 
to the phase-space integration and the unweighting of events.

The user of \aNLOs\ can create his/her own Sudakov pre-tabulation by
running {\tt gridsudgen.f} prior to any proper \aNLOs\ runs. For 
convenience, we shall include in the package a template of {\tt sudakov.f},
that should be appropriate for all applications to LHC physics (this relies 
on the numerical values on the r.h.s.'s of eqs.~(\ref{nodepar1}) 
and~(\ref{nodepar2})).

We conclude this appendix by returning to the problem of the actual
computation of the Sudakov values that constitute the entries of the
pre-tabulated grids. Any such entry is equal to 
\mbox{$\Delta_a^{(*)}\left(\qt;\QMt,\ell,\Mdips\right)$}
(see eq.~(\ref{sudscales})); in turn, at given particle identity
($a$), Sudakov type (that fixes $\ell$), and nodal points (that
fix $\qt$ and $\Mdips$), this is set equal to ${\cal P}$, the
no-emission probability as is computed directly by \PYe\ with the
specified conditions (note that this implies that, in the case
of a gluon, only one colour line, $\ell$, is considered). To this
end a new \PYe\ module, called {\tt pythia\_get\_no\_emission\_prob},
has been included in the code, whose return value is ${\cal P}$ and
that serves as a wrapper to call lower-lying (and mostly pre-existing) 
\PYe\ modules.

There are two crucial aspects in the computation of ${\cal P}$ that
must be stressed. We comment on each of them in turn.

Firstly, ${\cal P}$ is obtained by employing trial 
showering~\cite{Lonnblad:2001iq,Lonnblad:2012hz}. In such a procedure, the
parton shower, as is implemented in the event generator, is directly used to
sample the no-emission probability, in the following way. Given the two
values of the shower variable, $\QMt$ and $\qt$, a shower is generated, 
at the end of which one sets:
\beqn
{\cal P}_{\rm trial}&=&0\;\;\;\Longleftrightarrow\;\;\;
{\rm an~emission~with~}\qt<t<\QMt{\rm ~did~occur}\,,
\label{Ptr0}
\\
{\cal P}_{\rm trial}&=&1\;\;\;\Longleftrightarrow\;\;\;
{\rm an~emission~with~}\qt<t<\QMt{\rm ~did~not~occur}\,.
\label{Ptr1}
\eeqn
The procedure is repeated $N$ times, by only changing the random numbers
inherent to the showering. Note that the procedure is unitary: the number
of showers that result in ${\cal P}_{\rm trial}=0$, plus the number of those
leading to ${\cal P}_{\rm trial}=1$, is equal to $N$. One then defines:
\beq
{\cal P}=\frac{1}{N}\sum_{i=1}^N {\cal P}_{{\rm trial},i}\,,
\label{Pdef}
\eeq
with $i$ labelling the trial showers.

Secondly, usually the no-emission probability associated with showers
is {\em not} equal to $\Delta_a^{(*)}$ if $a$ is in the initial state.
For this to happen, the evolution needs to be done forwards rather than
backwards\footnote{The validity of some of the properties 
of backward evolution~\cite{Sjostrand:1985xi,Gottschalk:1986bk} which 
are usually assumed to hold has very recently been called into question
in ref.~\cite{Nagy:2020gjv}. It is beyond the scope of this work to
discuss the findings of the latter paper. We limit ourselves to 
observing that small differences, if any, between backward and forward
initial-state evolution are perfectly acceptable in the construction
of the $\Delta$ factor.}, as is typically the case in order not to lose 
efficiency given the constraint imposed by the value of the Bjorken $x$. 
Since such a constraint is irrelevant for the sake of pre-tabulation, 
a forward initial-state shower has been implemented, so that indeed
\mbox{$\Delta_a^{(*)}={\cal P}$} regardless of whether $a$ is in the
initial or final state.

The use of forward evolution for both initial and final-state showers 
is convenient for pre-tabulation, but poses a technical problem. Consider
the fact that any such shower is driven, among many other things, by
the form of the $z$ integral in eq.~(\ref{deltadef}). For reasons that
will soon become clear, we re-write that integral with a more generic
notation for the integration range. In the case of (anti)quark and gluon 
it reads, respectively, as follows:
\beqn
S_q &=&\int^{z^+}_{z^-} dz \Big[\hP_{qq}(z) + \hP_{gq}(z) \Big]\,,
\label{Sqdef}
\\
S_g &=&\int^{z^+}_{z^-} dz \Big[\hP_{gg}(z) + 
\sum_f\big(\hP_{q_fg}(z) + \hP_{\bar{q}_fg}(z)\big)\Big]\,,
\label{Sgdef}
\eeqn
with $z^\pm$ defined by phase-space and momentum-conservation
constraints specific to the given Sudakov type (roughly speaking,
one has \mbox{$z^+=1-\ep=1-\sqrt{t}/Q+\ord(t/\Qt)$}). As a matter 
of fact, in the case of the standard backwards initial-state showers $z^-$ 
is not relevant, because the lower bound of the integration range of 
the integrals playing the roles of $S_q$ and $S_g$ in backward evolution
is actually equal to the Bjorken $x$. Thus, for initial-state showers the 
definition of $z^-$ is simply not implemented in \PYe. Conversely, by 
evolving forwards a value $z^->0$ is necessary, in order to prevent the 
integrals of eqs.~(\ref{Sqdef}) and~(\ref{Sgdef}) from diverging owing to 
the $z\to 0$ soft-gluon singularity of $\hP_{gq}(z)$ and $\hP_{gg}(z)$. 
There are at least three different ways to address this issue, which
we now enumerate.
\begin{enumerate}
\item
As has been advocated in eq.~(\ref{deltadef}) and in the accompanying text,
set \mbox{$z^+=1-\ep$} and $z^-=\ep$, by analogy with the final-state
case, and because of symmetry considerations driven by the behaviour of
the Altarelli-Parisi kernels under a \mbox{$z\to 1-z$} transformation.
By explicit computation we obtain:
\beqn
&&\half\int^{1-\ep}_{\ep} dz \Big[\hP_{qq}(z) + \hP_{gq}(z) \Big]=
-2\CF\log\ep-\frac{3}{2}\,\CF+\ord(\ep)\,,\phantom{aaa}
\label{Sq1}
\\
&&\half\int^{1-\ep}_{\ep} dz \Big[\hP_{gg}(z) + 
\sum_f\big(\hP_{q_fg}(z) + \hP_{\bar{q}_fg}(z)\big)\Big]=
-2\CA\log\ep-\beta_0+\ord(\ep)\,,\phantom{aaaaaa}
\label{Sg1}
\eeqn
where $\beta_0$ is the first coefficient of the QCD $\beta$ function,
and the prefactor \mbox{$1/2=1/N_a$} takes eq.~(\ref{Naeq2}) into account.
\item
Set \mbox{$z^+=1-\ep$} and $z^-=1/2$. This is another way to
exploit the \mbox{$z\to 1-z$} properties of the kernels, which allows
one to map the $z=0$ soft singularity onto the $z=1$ one, with equivalent
physical meaning. Then:
\beqn
&&\int^{1-\ep}_{1/2} dz \Big[\hP_{qq}(z) + \hP_{gq}(z) \Big]=
-2\CF\log\ep-\frac{3}{2}\,\CF+\ord(\ep)\,,\phantom{aaa}
\label{Sq2}
\\
&&\int^{1-\ep}_{1/2} dz \Big[\hP_{gg}(z) + 
\sum_f\big(\hP_{q_fg}(z) + \hP_{\bar{q}_fg}(z)\big)\Big]=
-2\CA\log\ep-\beta_0+\ord(\ep)\,.\phantom{aaaaaa}
\label{Sg2}
\eeqn
Note that the $1/2$ prefactor that appears in eqs.~(\ref{Sq1}) 
and~(\ref{Sg1}) must not be used here, owing to the reduced integration 
range.
\item
Set \mbox{$z^+=1-\ep$} and $z^-=0$, and damp the $z=0$ soft singularity
by means of the formal replacements \mbox{$\hP_{ij}(z)\to z\hP_{ij}(z)$}.
We obtain:
\beqn
&&\int^{1-\ep}_{0} dz z\,\Big[\hP_{qq}(z) + \hP_{gq}(z) \Big]=
-2\CF\log\ep-\frac{3}{2}\,\CF+\ord(\ep)\,,\phantom{aaa}
\label{Sq3}
\\
&&\int^{1-\ep}_{0} dz z\,\Big[\hP_{gg}(z) + 
\sum_f\big(\hP_{q_fg}(z) + \hP_{\bar{q}_fg}(z)\big)\Big]=
-2\CA\log\ep-\beta_0+\ord(\ep)\,.\phantom{aaaaaa}
\label{Sg3}
\eeqn
Note that the multiplication by $z$ of the kernels effectively 
de-symmetrises the integrands, such that a symmetry factor equal 
to $1/2$ must not be employed.
\end{enumerate}
As the r.h.s.'s of eqs.~(\ref{Sq1})--(\ref{Sg3}) show, these strategies
are equivalent at the leading and subleading terms in $\ep$, and differ
only by terms of $\ord(\ep)$. We have implemented all of them, and indeed
have verified that the differences among their predictions are very small
(note that $\ep$ is typically a small parameter). Heuristically, we have 
observed that strategy \#3 leads to results which on average are the closest
to those of the standard backwards shower, and we have therefore adopted
it for our pre-tabulation.

We conclude this appendix by remarking that the use of trial showers,
being binary in nature (see eqs.~(\ref{Ptr0}) and~(\ref{Ptr1})) might
lead to a relatively coarse result for ${\cal P}$ defined as in 
eq.~(\ref{Pdef}). In order to alleviate this problem, we have employed
$N=10^4$ in our pre-tabulation. However, even that level of averaging may 
not be sufficiently smooth to be considered continuous for our purposes.
In view of this, we also make use of the shower-enhancement mechanism
described in ref.~\cite{Lonnblad:2012hz}. This stems from multiplying 
the splitting kernels by a factor $C$; such an artificial enhancement
has to be compensated at the level of no-emission probability --
eq.~(\ref{Ptr0}) becomes:
\beq
{\cal P}_{\rm trial}=\left(1-\frac{1}{C}\right)^n
\;\;\;\Longleftrightarrow\;\;\;
n\ge 1~{\rm emissions~did~occur~in~the~}(\qt,\QMt){\rm ~range}\,,
\label{Ptr02}
\eeq
while eq.~(\ref{Ptr1}) is unchanged. One can then still employ
eq.~(\ref{Pdef}).
The fine-graining of the enhanced trial-shower procedure is particularly 
important for values of $q^2$ that would result in a high emission rate, 
i.e.~in the soft/collinear regions, since there the Sudakov is a 
fastly-varying function. To this end, we have observed that $C\ge 3$
is sufficient for the applications we have considered in this paper.

\phantomsection
\addcontentsline{toc}{section}{References}
\bibliographystyle{JHEP}
\bibliography{mcdelta}

\end{document}